\documentclass[aps,prb,twocolumn,amsmath,amssymb,floatfix,eqsecnum]{revtex4-1}

\usepackage{graphicx}
\usepackage{dcolumn}
\usepackage{bm}
\usepackage{hyperref}
\usepackage{braket}
\usepackage{subfigure}

\begin{document}
\newcommand{\ra}{\rangle}
\newcommand{\la}{\langle}
\newcommand{\dg}{{\dagger}}
\newcommand{\pdg}{{\phantom\dagger}}
\newcommand*\conj[1]{\overline{#1}}

\title{Chern-Simons Composite Fermion Theory of Fractional Chern Insulators}

\author{Ramanjit Sohal}
\author{Luiz H. Santos}
\author{Eduardo Fradkin}
\affiliation{Department of Physics and Institute for Condensed Matter Theory, University of Illinois at Urbana-Champaign, 1110 West Green Street, Urbana, Illinois 61801-3080, USA}

\date{\today}

\begin{abstract}
We formulate a Chern-Simons composite fermion theory for Fractional Chern Insulators (FCIs), whereby bare fermions are mapped into composite fermions coupled to a lattice Chern-Simons gauge theory. We apply this construction to a Chern insulator model on the kagome lattice and identify a rich structure of gapped topological phases characterized by fractionalized excitations including states with unequal filling and Hall conductance. Gapped states with the same Hall conductance at different filling fractions are  characterized as realizing distinct symmetry fractionalization classes. 
\end{abstract}

\maketitle

\section{Introduction} Recent years have seen increasing interest in Fractional Chern Insulators (FCIs), materials which realize analogues of the fractional quantum Hall effect (FQHE) on lattices in the absence of a net magnetic field. Numerical studies have shown that FCIs can be realized by partially filling a band with non-trivial Chern number.\cite{Neupert11,Sheng11,Tang11,Sun11,Regnault11,Liu-2012,Wu12,Lauchli-2013,Liu13} Encouragingly, the observation of lattice FQH states was recently reported in a bilayer graphene/hexagonal boron nitride heterostructure.\cite{Spanton17}

Despite the striking similarities between FCIs and FQHE states, the differences between the physical mechanisms giving rise to these two states of matter render a theoretical description of FCIs a challenging endeavor.
In semiconductor heterostructures where the FQHE is traditionally studied, a two-dimensional electron gas (2DEG)
is subject to a uniform magnetic field of the order 10T, which is externally applied
perpendicularly to the 2DEG. For magnetic fields of this order, 
the magnetic flux per unit cell of the underlying periodic lattice potential experienced by the electrons
is much smaller than the fundamental flux quantum $\hbar c/e$, or equivalently, the magnetic length 
$\ell_{B} = \sqrt{\hbar c/eB}$
(where $h$ is the Planck's constant, $c$ the speed of light, $e$ the electron's charge and $B$ the magnetic field)
is much larger than the lattice constant of the underlying periodic lattice potential experienced by the electrons, 
which justifies neglecting lattice effects. Under these circumstances, one treats electrons in a continuum approximation
where the magnetic field organizes the single particle energy states into highly degenerate Landau levels.
The special mathematical properties associated with the single particle energy states of the lowest Landau level, in particular,
have been largely explored towards a theoretical understanding of the essential properties of the FQH liquid,
such as the existence of quasiparticles with fractional charge and statistics.
\cite{Laughlin83,Arovas84,Halperin84,Moore-1991}

In FCIs, on the other hand, the single particle states constitute a topologically non-trivial Chern band.\cite{Thouless82}
When completely filled, a Chern band with Chern number $C_0 = 1$ yields a Hall conductance $e^{2}/h$ exactly
as a completely filled Landau level does. Nevertheless, unlike the flat energy Landau levels, 
Chern bands have a 
a non-zero bandwidth, which renders the analytical treatment of FQH states supported in partially
filled Chern bands much more challenging when compared to the partially filled Landau level situation.
As such, much of the understanding about the topological properties of FCIS has relied on numerical
methods such as exact diagonalization (ED) and density matrix renormalization group (DMRG).
\cite{Neupert11,Sheng11,Tang11,Sun11,Regnault11,Liu-2012,Wu12,Lauchli-2013,Liu13}

The development of an analytical description of FCI states is thus of great interest. Although FCIs exhibit characteristics similar to the continuum FQHE, lattice effects are expected to give rise to rich new physics. Even in the continuum FQHE, the presence of a square lattice potential yields states with Hall conductance unequal to the filling factor.\cite{Kol93} FCIs are additionally expected to support non-Abelian states resulting from partial filling of Chern bands with $|C_0|>1$, \cite{Lu12,Sterdyniak13} non-Abelian defects, \cite{Barkeshli12} and anyonic excitations with  fractionalized symmetry quantum numbers, thus exhibiting `symmetry fractionalization'.\cite{Essin13, Chen17} Although several approaches have been used to study FCIs, an analytic description starting from a microscopic theory is still lacking.

There are two appealing motivations for pursuing this analytic description. 
Firstly, given the topological nature encoded in the ground state and low-energy excitations of FCIs, 
one hopes to uncover the essential aspects underlying the mechanism of electron fractionalization,
without having to rely intensively on details of the systems, as one does in ED and DMRG studies.
Secondly, the analytical approach we pursue in this work - as shall be presented in detail later -
predicts candidate FCI states stabilized by local interactions in Chern bands in a series of partial filling fractions, 
whose properties can be targeted in future ED and DMRG studies as well as in experiments.
Thus our work provides a bridge between analytic, numerical and experimental studies.

The aim of this paper is to  provide an analytical description of FCI states on a specific kagome lattice model in terms of a Chern-Simons composite fermion (CF) theory. The CF approach has been used successfully in describing the conventional FQHE.\cite{Jain89,Lopez91} Heuristically, in this picture the electrons nucleate fluxes of an emergent Chern-Simons gauge field such that at the mean field level, the CFs fill an integer number of Landau levels in the new effective flux. The IQH states of the CFs then correspond to FQH states of the original fermions. Quantum fluctuations about these IQH states endow the excitations with their correct fractionalized quantum numbers. Numerical studies suggest that a CF picture is also relevant for the lattice FQHE \cite{Moller09,Moller15} and so may be applicable to FCIs,
as is also suggested by recent analytical work,\cite{Murthy12,Murthy14} providing further motivation for our study. The methodology we use here is similar to that used in the continuum, but with a local and gauge-invariant  Chern-Simons action on the lattice.\cite{Fradkin89,Eliezer92,Kumar14,Sun15} This is necessary as we work in the tight binding limit; in contrast, previous studies of the lattice FQHE \cite{Kol93} considered the opposite limit in which Landau levels are dressed by a lattice potential.

This paper is organized as follows. We begin by outlining in Sec. \ref{sec:model} the Chern insulator model on the kagome lattice to which we apply our analysis. Next, in Sec. \ref{sec:flux-attachment}, we review the CF mapping and the consistent definition of a Chern-Simons action on the kagome lattice as discussed in Refs. [\onlinecite{Kumar14}] and [\onlinecite{Sun15}] (see Sec. \ref{sec:LCS}). Special care will be given  
to the boundary conditions, focusing on the case of the torus. In Sec. \ref{sec: Mean Field Theory} we construct gapped states of the CFs arising at the mean field level (i.e. the average field approximation) which are identified as candidate FCI states. The construction of the mean-field theory requires the self-consistent derivation of a class of Hofstadter states. 
In Sec. \ref{sec:TQFT} we derive the effective topological field theory describing the gapped low-energy states, from which we can extract the Hall conductance, fractional excitations, and ground state degeneracy. In Sec. \ref{sec:hofstadter} we use the results of the previous sections to identify candidate FCI states from the composite fermion Hofstadter spectrum and characterize their topological properties. It is found that there are two classes of FCI states which can be realized: those with Hall conductance equal to the filling factor and those for which these two quantities are unequal. We then discuss how FCI states with the same topological order arising at different filling fractions can be viewed as realizing distinct symmetry fractionalization classes which are derived in Sec.\ref{sec:symmetry-fractionalization}. We provide concrete predictions arising from this classification which may be verified via numerical studies. Before concluding, we offer some remarks in Sec. \ref{sec:FCI vs FQHE} about the distinctions between FCIs and lattice FQHE states. Sec. \ref{sec:conclusions} is devoted to our conclusions. 

\section{Model} 
\label{sec:model}

The model we consider is a Chern insulator of spinless fermions with nearest-neighbor hopping on the kagome lattice, subject to a staggered magnetic field with net zero flux per unit cell.\cite{Green10} 
The Hamiltonian of the model is
\begin{align}
H=&J\sum_{\la {\bm x}, {\bm x}' \ra} \left[\psi^\dg({\bm x})e^{i  \phi({\bm x},{\bm x}')} \psi({\bm x}') + h.c. \right]\nonumber\\
&+ \frac{g}{2} \sum_{{\bm x},{\bm y}} n({\bm x}) V({\bm x}-{\bm y}) n({\bm y}) \label{eqn:model}
\end{align}
Here $J$ is the hopping amplitude on the kagome lattice, $g$ is the coupling constant, and $V({\bm x}-{\bm y})$ is the interaction between the fermions densities, the occupation numbers  labelled by $n({\bm x})$. In this paper we will assume that the interaction is for nearest-neighbors on the kagome lattice, but it can be a general interaction as well.

\begin{figure}[hbt]
  \includegraphics[width=0.37\textwidth]{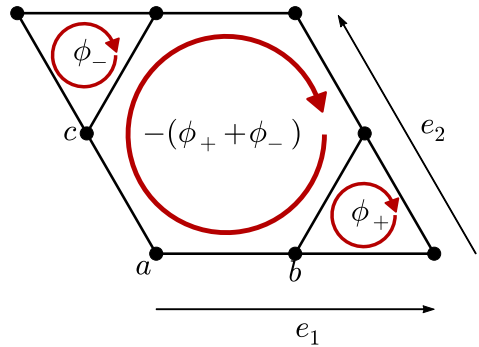}
  \caption{(Color online) Kagome lattice unit cell with phases arising from a magnetic flux indicated by arrows. The net flux is zero.}
  \label{fig:unitcell}
\end{figure}

The staggered magnetic field is represented by the fixed phase field $\phi({\bm x},{\bm x}')$, defined on the links of the kagome lattice,
such that the  flux of this field is $\phi_+$ ($\phi_-$) into the page through the up (down) triangles of the lattice as shown in Fig \ref{fig:unitcell}.
Also illustrated in Fig. \ref{fig:unitcell} are the sublattice structure and definition of the lattice vectors ${\bm e}_{1,2}$. We will focus on the case of $\phi_\pm = \pi/2$ which preserves the lattice point-group symmetry and yields three well-separated bands with the lowest band possessing Chern number $C_0 = +1$. 
The band structure is discussed further in Appendix \ref{sec:non-interacting band stucture}. We are interested in gapped topological phases arising from partially filling this lowest band; the fraction of it which is filled will be denoted by $n_L$.

In general, the formation of an FCI requires that the bandwidth of the partially filled Chern band are at most of the order of the band gaps of the non-interacting problem. Hence, the spinless fermions in the partially filled band can be regarded as being strongly correlated. Without this condition the system would be a metal, and not an FCI. In the specific staggered flux model we chose, the bandwidth is comparable to the gap between the band. It is, however, still possible to stabilize an FCI, in principle, if interactions are strong enough.

\section{Flux Attachment} 
\label{sec:flux-attachment}

The mapping of fermions to CFs is accomplished via a mapping onto an equivalent system  of spinless fermions, the composite fermions, minimally coupled to a dynamical lattice Chern-Simons gauge field with a coupling parameter that we will denote by $\theta$. We use the lattice Chern-Simons gauge theory on a kagome lattice on a 2D torus, as defined in Ref.[\onlinecite{Sun15}]  (a generalization of the approach of Refs.[\onlinecite{Fradkin89,Eliezer92}]). This is an \textit{exact} mapping provided the Chern-Simons theory can be well-defined on the lattice in a gauge-invariant fashion. That this is possible on a large class of lattices (including the kagome lattice) was shown in Ref. [\onlinecite{Sun15}]. After performing this transformation our system is described by the action
\begin{equation}
S[\psi,\psi^\dg,A_\mu,B_\mu] = S_F[\psi,\psi^\dg,A_\mu] + S_{int}[A_\mu] + S_{CS}[A_\mu, B_\mu]
\end{equation}
where $\psi$ is the CF field, $A_\mu$ is the statistical gauge field, and $B_\mu$ is a hydrodynamic gauge field required for  flux attachment to be defined consistently on a torus 
as described below.\cite{Fradkin2013} Here $S_F$ is the fermionic action, $S_{int}$ is a fermion density-density interaction (as will be shown, flux attachment allows one to make a formal substitution of the densities with fluxes), and $S_{CS}$ is the Chern-Simons action. In the following we will discuss in detail the actions $S_F$, $S_{CS}$, and $S_{int}$ in turn.

The fermionic part of the action is of the usual form,
\begin{align}
\begin{split}
&S_F[\psi,\psi^\dg,A_\mu] = \int_t \sum_{{\bm x}} \psi^\dagger ({\bm x},t)iD_0 \psi({\bm x},t) \\
 &- \int_t J \sum_{\la {\bm x}, {\bm x}' \ra} (\psi^\dg({\bm x},t)e^{i (A_j({\bm x},t) + \phi({\bm x},{\bm x}'))} \psi({\bm x}',t) + h.c. ),
\end{split}
\end{align}
where $D_0 = \partial_0 + iA_0$ is the covariant time derivative and $\la {\bm x}, {\bm x}' \ra$ indicates a sum over nearest neighbors. The temporal components of the gauge field, $A_0({\bm x})$, live on the sites of the lattice whereas the spatial components, $A_i({\bm x})$ ($i = 1, \dots, 6$), live on the links. 

The form of the lattice Chern-Simons action is less intuitive. For a more detailed construction for a large class of lattices we direct the reader to Ref. [\onlinecite{Sun15}]. In the following two subsections we will describe first the lattice formulation of the action for the statistical gauge field $A_\mu$ and subsequently the lattice formulation of an equivalent theory involving both $A_\mu$ and the hydrodynamic field $B_\mu$ which can can be defined on topologically non-trivial manifolds.

\subsection{Lattice Chern-Simons}
\label{sec:LCS}

Broadly speaking, a theory of fermions coupled to a Chern-Simons field is a theory of flux-charge composites. Hence a Chern-Simons action must enforce a Gauss' Law constraint which attaches fluxes to the matter fields in addition to being gauge invariant. On the lattice, the fermions reside on the lattice sites and so it is natural to define the fluxes to live in the plaquettes (i.e. the sites of the dual lattice). So, in order to be able to consistently define a flux attachment condition, the lattice must be such that one can uniquely associate each site to a plaquette. This is indeed the case for the kagome lattice as well as the dice and square lattices but not, for instance, the honeycomb or triangular lattices.

\begin{figure}[!htb]
  \subfigure[]{
     \includegraphics[width=0.22\textwidth]{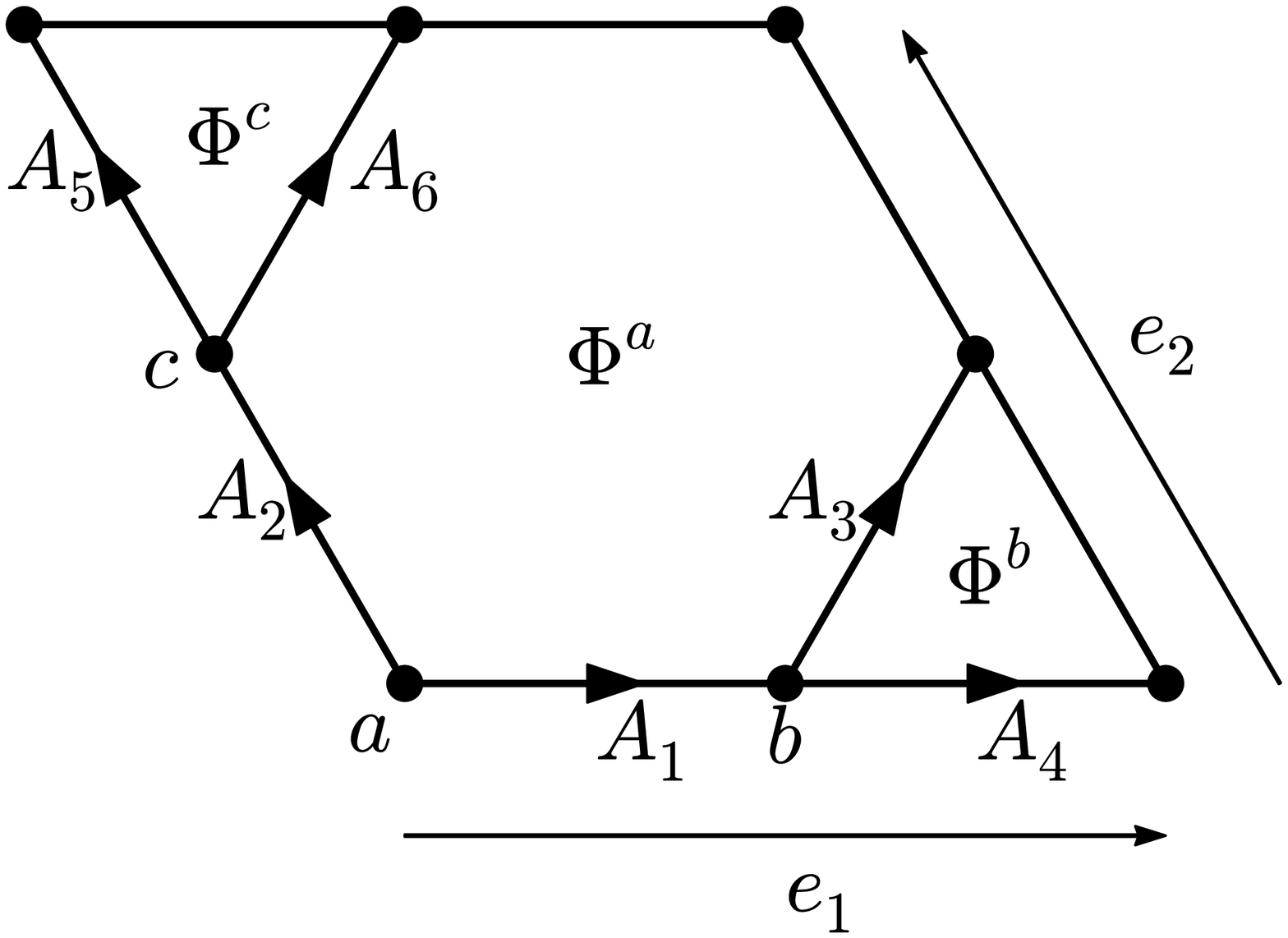} \label{fig:CSunitcell-a}
  }
  \subfigure[]{
     \includegraphics[width=0.22\textwidth]{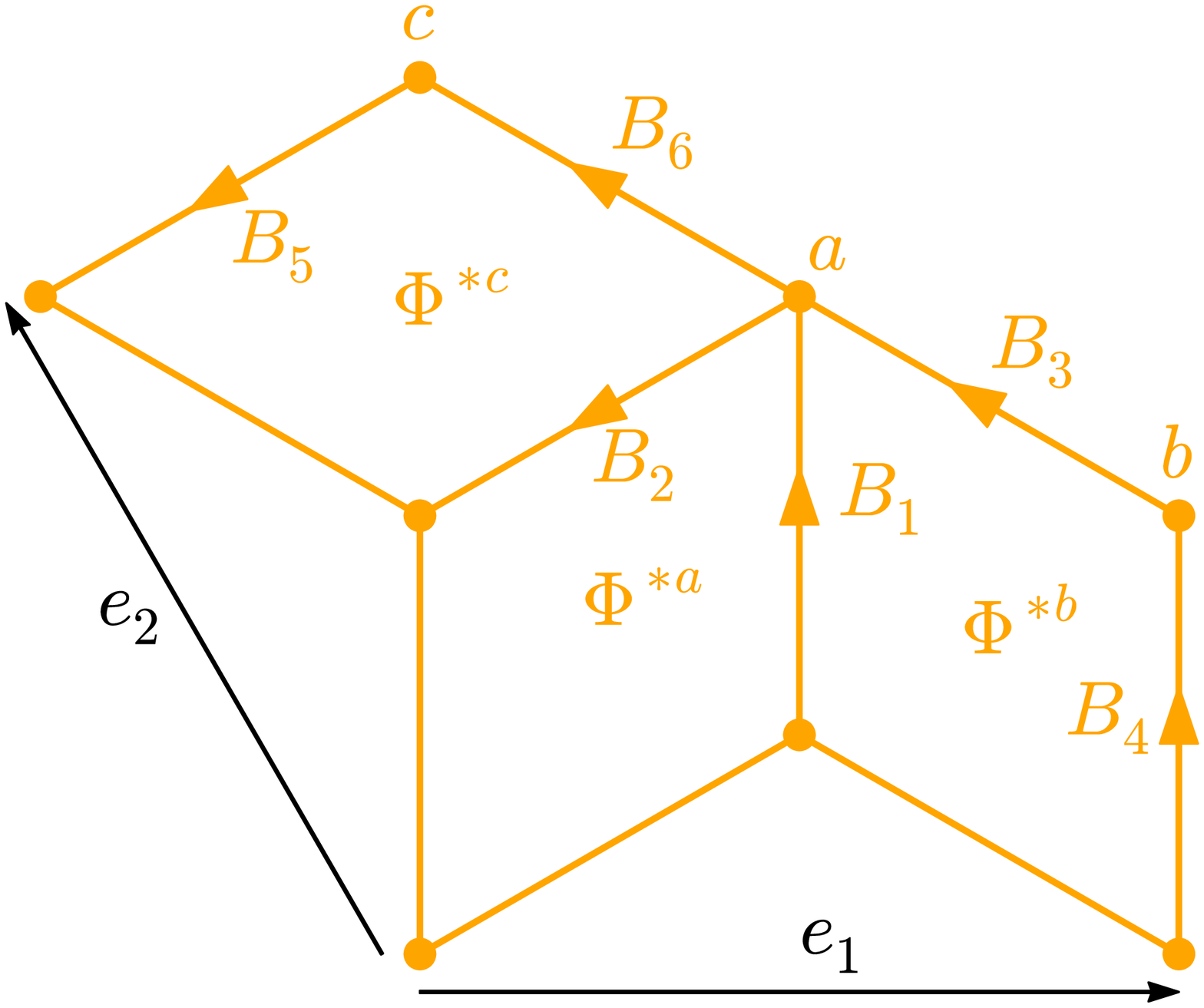} \label{fig:CSunitcell-b}
  }
  \subfigure[]{
     \includegraphics[width=0.22\textwidth]{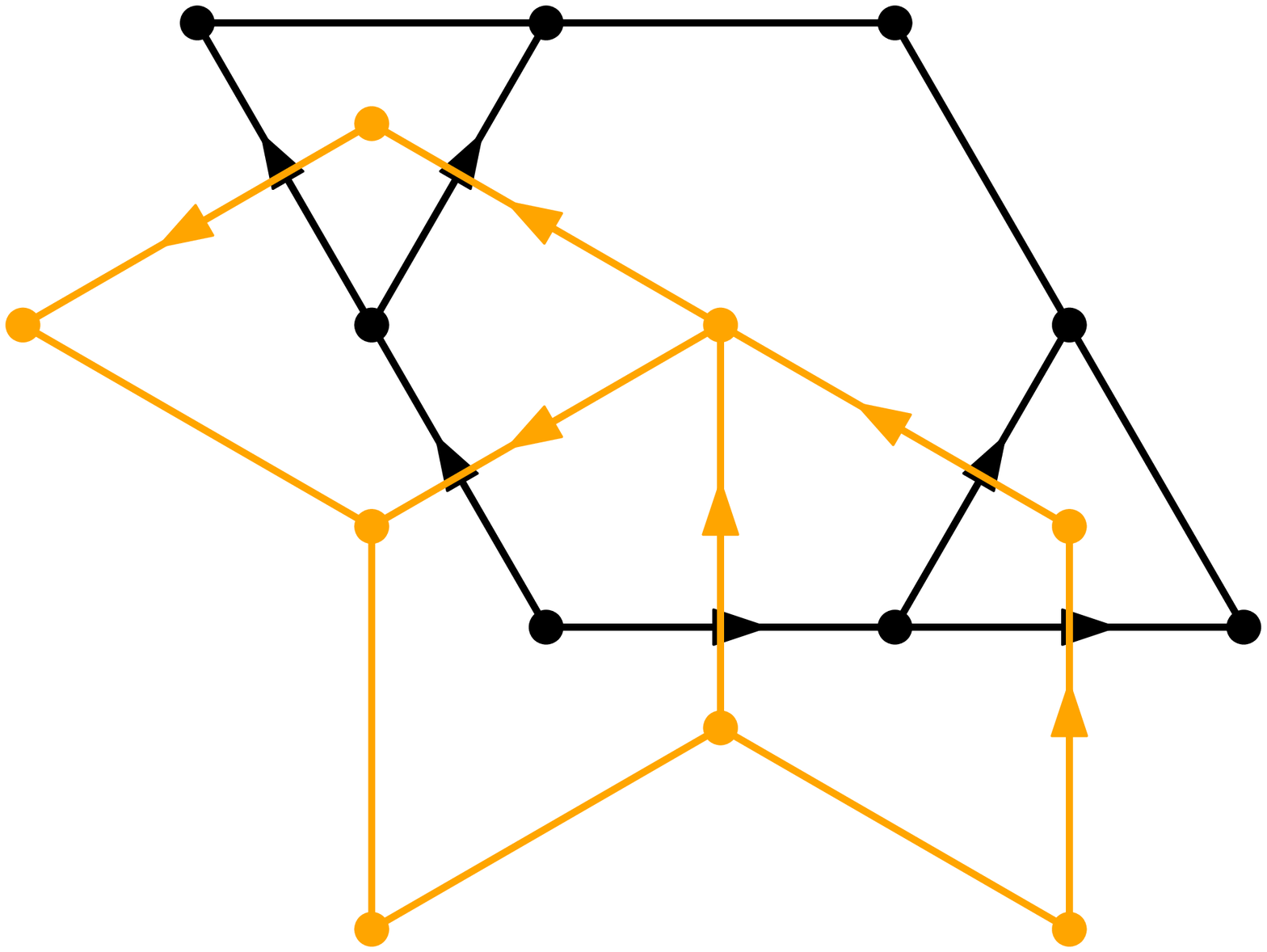} \label{fig:CSunitcell-c}
  }
  \caption{(Color online) a) Kagome lattice unit cell with spatial components of the statistical gauge field $A_i$ and fluxes $\Phi^\alpha = J^\alpha_i A_i$. b) Dice lattice unit cell with the hydrodynamic gauge field $B_i$ and fluxes $\Phi^{*\alpha} = J_i^{*\alpha} B_i$. c) Orientation of the dual (dice) lattice relative to the direct (kagome) lattice.}
  \label{fig:CSunitcell}
\end{figure}

As shown in Ref. [\onlinecite{Kumar14}] and [\onlinecite{Sun15}], it follows from the above considerations that one consistent formulation for a Chern-Simons action on the kagome lattice is given by \footnote{Note that what we call $\mathcal{M}_{ij}$ is the $K_{ij}$ matrix of Refs. [\onlinecite{Eliezer92,Kumar14,Sun15}]; this is to avoid confusion with the $K$-matrix of Eq. \ref{eqn:KMatrixCtmAction}. Additionally, as we have written the action, $J_i^\alpha$ and $\mathcal{M}_{ij}$ are operators while in Ref. [\onlinecite{Kumar14}] these objects are expressed as the functions $J_i({\bm x}-{\bm y})$ and $\mathcal{M}_{ij}({\bm x}-{\bm y})$ with the action containing a sum over ${\bm x}$ and ${\bm y}$. This is a matter of notation -- in this and the following sections we will use one or the other convention depending on which is most convenient.}
\begin{align}
\begin{split}
S_{CS}[A_\mu] = \theta \int dt \sum_{{\bm x}} & \left[  A^\alpha_0 ({\bm x},t) J^\alpha_i A_i({\bm x},t) \phantom{\frac{1}{2}}\right. \\ 
&\left. - \frac{1}{2} A_i({\bm x},t) \mathcal{M}_{ij} \partial_t{A}_j({\bm x},t) \right].
\end{split} \label{eqn:discCS}
\end{align}
Here the sum is over unit cell positions and the index $\alpha=a,b,c$ denotes the sublattice. 
As noted above, the temporal components of the gauge field, $A_0^\alpha$, live on the lattice sites whereas the spatial components, $A_i$, live on the links. The orientation of the spatial gauge fields is shown in Fig. \ref{fig:CSunitcell-a}. In the context of flux attachment the coupling $\theta$ is given by $\theta = 1 / 2\pi (2k), \, k \in \mathbb{Z}$. Now, the first term in the action enforces the flux-attachment (or Gauss' Law) constraint
\begin{equation}
\label{eqn:fluxattach}
n^\alpha({\bm x}) = \theta J_i^{\alpha} A_i({\bm x}) \equiv \theta \Phi^\alpha({\bm x}).
\end{equation}
The fluxes $\Phi^\alpha ({\bm x}) = J_i^{\alpha} A_i({\bm x})$ live in the kagome lattice plaquettes as shown in Fig. \ref{fig:CSunitcell-a}. The $J_i$ vectors may be viewed as discretized curl operators on the kagome lattice,
\begin{align}
\begin{split}
J^a &= (1, -1, 1, -s_2, -s_1, -1), \\
J^b &= (0, s_1, -1, 1, 0, 0), \\
J^c &= (-s_2, 0, 0, 0, -1, 1),
\end{split}
\end{align}
where we have used the lattice shift operators $s_i$ which are defined as $s_i f({\bm x}) = f({\bm x} + {\bm e}_i)$. Hence the Gauss law ties the statistical fluxes in the hexagon, up triangle, and down triangle to the $a$, $b$, and $c$ sites, respectively.

Note that the assignment of fluxes to sites necessarily breaks the rotational symmetry of the lattice. This will be true for any choice of assignment and so the lattice point-group symmetry is not respected by the lattice Chern-Simons formulation we have chosen. That being said, the mapping of fermions to composite fermions is an exact one at the level of the path integral and so the ground state predicted by our theory (at the full quantum level) should respect the lattice symmetries, provided there is no spontaneous symmetry breaking. We will return to this issue in the discussion of our mean field theory analysis.

The second term in the Chern-Simons action of Eq.\eqref{eqn:discCS} enforces the commutation relations. The matrix kernel $\mathcal{M}_{ij}$ must be chosen so as to make the theory gauge-invariant. This ensures that the Gauss' Law constraint can be applied consistently on different plaquettes, in other words, $[\Phi^\alpha({\bm x}),\Phi^{\alpha'}({\bm y})] = 0$. It was found previously \cite{Kumar14} that on the kagome lattice the matrix kernel has the form
\begin{widetext}
\begin{align}
\label{eqn: kagome matrix kernel}
\mathcal{M}_{ij} = \frac{1}{2}\left(
 \begin{matrix}
  0 & -1 & 1 & -s_2 & s_1 + s_2^{-1} & -1 + s_2^{-1} \\
  1 & 0 & 1-s_1^{-1} & -s_2-s_1^{-1} & s_1 & -1 \\
  -1 & s_1-1 & 0 & 1-s_2 & s_1 & -1 \\
  s_2^{-1} & s_1+s_2^{-1} & s_2^{-1}-1 & 0 & s_1s_2^{-1} & s_2^{-1} \\
  -s_2-s_1^{-1} & -s_1^{-1} & -s_1^{-1} & -s_2 s_1^{-1} & 0 & 1-s_1^{-1} \\
  1-s_2 & 1 & 1 & -s_2 & s_1 - 1 & 0
 \end{matrix}
\right).
\end{align}
\end{widetext}
The standard canonical quantization procedure then yields $[A_i({\bm x}),A_j({\bm y})] = -\frac{i}{\theta}\mathcal{M}^{-1}_{ij}({\bm x}-{\bm y})$ so that only neighboring gauge fields have non-trivial commutation relations. 
(See Appendix~\ref{sec:spectrum of K} for a discussion of the spectrum of $\mathcal{M}_{ij}$.) This provides a fully consistent definition of a local Chern-Simons action on the kagome lattice with trivial topology.

\subsection{Lattice Chern-Simons on a Torus}
\label{sec:LCS-torus}

As noted above, when performing our mapping, the Chern-Simons parameter is given by $\theta = 1/2\pi(2k)$. This is the form of the flux attachment used in Ref. [\onlinecite{Kumar14}]. However, the coefficient $\theta$ is not properly quantized and so the theory cannot be defined on closed manifolds with non-zero genus.\cite{Witten89} Much as in the continuum case,\cite{Lopez99,Fradkin2013} this problem is resolved by introducing the auxiliary, hydrodynamic field $B_\mu$ which lives on the dual lattice which, in our case, is  the \textit{dice lattice}. In particular, the temporal components, $B_0^\alpha$, live on the sites of the dice lattice whereas the spatial components, $B_k$, live on the links. Following the conventions of Ref. [\onlinecite{Sun15}], the orientations of the $B_k$ are obtained by rotating the arrows on the direct lattice counter-clockwise until they align with the links of the dual lattice. Fig. \ref{fig:CSunitcell-b} shows the definition of the spatial components of the $B_\mu$ field while Fig. \ref{fig:CSunitcell-c} illustrates the relative orientation of the direct and dual lattices. The resulting action is given by
\begin{widetext}
\begin{align}
\begin{split}
S_{CS}[A_\mu, B_\mu] &= -\frac{2k}{2\pi} \int dt \sum_{\bm x} B_0^\alpha ({\bm x},t) J^{*\alpha}_i B_i({\bm x},t) - \frac{1}{2} B_i({\bm x},t) \mathcal{M}^*_{ij} \dot{B}_j ({\bm x},t) \\
&+ \frac{1}{2\pi} \int dt \sum_{\bm x} A_0^\alpha ({\bm x},t) J_i^{*\alpha} B_i ({\bm x},t) + B_0^\alpha ({\bm x},t) J_i^\alpha A_i ({\bm x},t) - \dot{B}_j ({\bm x},t) A_j({\bm x},t).
\end{split} \label{eqn:dualCS}
\end{align}
\end{widetext}
The first two terms give the Chern-Simons action for $B_\mu$ on the dual lattice whereas the remaining three terms give the `BF' coupling between the $A_\mu$ and $B_\mu$ fields. This is the discretized form of the continuum action
\begin{align}
\mathcal{L}_{CS}^{\mathrm{ctm}}[A_\mu, B_\mu] =  -\frac{2k}{2\pi} \epsilon^{\mu\nu\lambda} B_\mu \partial_\nu B_\lambda + \frac{1}{2\pi} \epsilon^{\mu\nu\lambda} A_\mu \partial_\nu B_\lambda.
\label{eqn:ctmDualCS}
\end{align}
known as the BF theory.\cite{Horowitz-1989}
The objects $J^{*\alpha}_i$ and $\mathcal{M}^*_{ij}$ are the analogues of $J^{\alpha}_i$ and $\mathcal{M}_{ij}$ on the dice lattice. Explicitly, we have
\begin{align}
\begin{split}
J^{*a} &= (1, 1, 0, -s_1^{-1}, -s_2^{-1}, 0), \\
J^{*b} &= (-1, 0, 1, 1, 0, -s_2^{-1}), \\
J^{*c} &= (0, -1, -s_1^{-1}, 0, 1, 1)
\end{split}
\end{align}
so that the $B_\mu$ fluxes are given by $\Phi^{*\alpha}({\bm x}) = J^{*\alpha}_i B_i({\bm x})$. Using the construction of Ref. [\onlinecite{Sun15}], we find
\begin{widetext}
\begin{align}
\mathcal{M}^*_{ij} = \frac{1}{2}\left(
 \begin{matrix}
  0 & -1 & 1 & -1+s_1^{-1} & s_2^{-1} & s_2^{-1} \\
  1 & 0 & -s_1^{-1} & -s_1^{-1} & 1-s_2^{-1} & -1 \\
  -1 & s_1 & 0 & 1 & s_1 & -s_1-s_2^{-1} \\
  1-s_1 & s_1 & -1 & 0 & s_1s_2^{-1} & s_2^{-1} \\
  -s_2 & -1+s_2 & -s_1^{-1} & -s_2 s_1^{-1} & 0 & 1 \\
  -s_2 & 1 & s_2 + s_1^{-1} & -s_2 & -1 & 0
 \end{matrix}
\right).
\end{align}
\end{widetext}
It can be seen that $\mathcal{M}^* = - \mathcal{M}^{-1}$ and so the $\mathcal{M}$ matrices are non-singular. Similar to the calculation shown in Appendix E of Ref. [\onlinecite{Sun15}], one can integrate out $B_\mu$ in Eq. (\ref{eqn:dualCS}) to recover Eq. (\ref{eqn:discCS}) and so the two theories are formally equivalent. However, Eq. (\ref{eqn:dualCS}) has properly quantized coefficients while Eq. (\ref{eqn:discCS}) does not and so the former is well-defined on topologically non-trivial manifolds whereas the latter is not. In using Eq. (\ref{eqn:dualCS}) our flux attachment procedure can be performed on a toroidal geometry and so we will be able to safely infer the topological field theory describing our FCI states.

Having completed our description of the lattice Chern Simons action, we return to the flux attachment procedure. Due to their coupling to the statistical gauge field, the $\psi$ fields have statistical angle $\delta = 2\pi k$ relative to the original fermionic statistics
 (as per the flux attachment constraint), and so by choosing the Chern Simons parameter $k$ to be an integer, we ensure that the CFs are indeed fermions. The $2\pi$ periodicity of the statistical angle implies that theories with different integral values of $k$ should be equivalent. This property is broken at the mean-field level but is recovered at the quantum level.

Lastly, $S_{int}$ is assumed to describe a density-density interaction which, due to the flux attachment constraint 
Eq. \ref{eqn:fluxattach}, has the form 
\begin{align}
S_{int}[A_\mu] = -\frac{1}{2}\int_t \sum_{{\bm x},{\bm y}} \frac{g}{16\pi^2 k^2}\Phi^{\alpha}({\bm x},t) V({\bm x}-{\bm y})_{\alpha \beta} \Phi^{\beta}({\bm y},t) \label{eqn:S_int}
\end{align}
where the explicit sum is over unit cell positions and there is an implicit sum over the sublattices, $\alpha,\beta = a,b,c$.
In this form, the interaction term does not enter in the fermionic part of the action. Instead, it is a parity-even contribution to the action of the gauge fields, similar in this sense to a Maxwell term for the fluxes. Since it involves more derivatives than the Chern-Simons term it is irrelevant as far as the long-wavelength fluctuations are concerned. However, it affects the local energetics and it enters in the mean-field equations (as we will see below). This analysis very much parallels what is done for the FQH states in the continuum, e.g. see Ref. [\onlinecite{Lopez91}], whose strategy we will follow closely.
Hence, we will first identify the gapped FCI ground states, in the present formalism, with gapped composite 
fermion ground states, which will be analyzed in mean field theory in Sec. \ref{sec: Mean Field Theory}.
After integrating out the gapped composite fermions, the universal contribution of the topological 
theory is encoded in an effective action $\mathcal{L}_{\mathrm{eff}}[A_\mu,B_\mu]$ containing Chern-Simons 
terms [see Eq.\eqref{eqn: effective action}]. 
In Appendix \ref{sec:MFT_interactions} we check for two examples of gapped states that, at the mean field level, the gap persists for a range of interactions strengths. Thus, the states that we will identify will not have any (infrared) instabilities since it is protected by the gap. In fact, for large enough interactions there should be a phase transition to a state with a broken translation symmetry. 
Nevertheless, since $S_{int}$ is quadratic in fluxes, it is irrelevant relative to 
the Chern-Simons action of Eq. \eqref{eqn: effective action}, and its presence does not affect the universal properties of the topological fixed points of the theory, provided the gap has not closed.

\section{Mean Field Theory} 
\label{sec: Mean Field Theory}

Analogous to the approach taken for the CF theory of the conventional FQHE, we wish to identify gapped states of the CFs at the mean field level. These will correspond to candidate FCI states.
The specific form of the interaction will determine whether these  states remain gapped and if they are energetically favorable to other potential ground states.

We first note that after the Jordan-Wigner mapping the action has become quadratic in the fermions and so they may formally be integrated out, yielding the effective action
\begin{align}
S_{\mathrm{eff}}[\psi,\psi^\dg,A_\mu,B_\mu] = -i\log \mathrm{tr} \left[iD_0-H(A)\right] + S_{CS}[A_\mu, B_\mu]
\end{align}
where $H(A)$ is the Hamiltonian describing fermions hopping on the lattice subject to the original magnetic fluxes as well as the statistical gauge field. The mean field ground states are solutions to the saddle point equations which are obtained by extremizing the effective action with respect to the gauge fields. We obtain
\begin{align}
\la n^\alpha ({\bm x})\ra &= \frac{1}{2\pi(2k)} \sum_{{\bm y}} J_i^\alpha ({\bm x}-{\bm y}) A_i({\bm y}) = \theta \la \Phi^\alpha ({\bm x}) \ra \label{eqn:MF1} \\
\begin{split}
\la j_k ({\bm x}) \ra &= \theta \sum_{{\bm y}}\left[ A^\alpha_0 ({\bm y},t) J^\alpha_k({\bm y}-{\bm x}) - \mathcal{M}_{kj}({\bm x}-{\bm y}) \dot{A}_j({\bm y},t) \right] \\ 
		& - g \theta^2 \sum_{\bm y, \bm z} J_k^\alpha (\bm z - \bm x) V_{\alpha\beta}(\bm z - \bm y ) B^\beta( \bm y )
\end{split} \label{eqn:MF2}
\end{align}
where $\la n^\alpha({\bm x}) \ra$ and $\la j_k ({\bm x}) \ra$ are the fermion density on sublattice $\alpha$ and current on link $k$, respectively. Explicitly,
\begin{align}
\la n^\alpha ({\bm x})\ra = - \frac{\delta S}{\delta A_0^\alpha ({\bm x})}, \quad \la j_k ({\bm x}) \ra = -\frac{\delta S}{\delta A_k ({\bm x})}.
\end{align}
Note that Eq. (\ref{eqn:MF1}) is simply the flux attachment constraint imposed on average. 
Now, we are interested in time-independent solutions which preserve the translational symmetry of the lattice (i.e. $\la n^\alpha({\bm x}) \ra = \la n^\alpha({\bm x}+{\bm e}_i) \ra$ for $i =1,2 $ and likewise for all other gauge-invariant quantities). In addition we assume that the currents on all links are equal in magnitude and circulate with the same chirality around both up and down triangular plaquettes so that $j \equiv \la j_{1,3,5} \ra = -\la j_{2,4,6} \ra$. 
Since generically $j \neq 0$, we see from Eq. (\ref{eqn:MF2}) that the $A_0^\alpha$'s can be different from one another, giving rise to unequal fermion densities on the three sublattices. So, given this ansatz, the resulting mean field equations are satisfied by
\begin{align}
\begin{split}
n^a &= \theta \Phi^a = n_L/3 - 2\Delta, \\
n^b &= \theta \Phi^b = n_L/3 + \Delta, \\
n^c &= \theta \Phi^c = n_L/3 + \Delta, \label{eqn:densityansatz}
\end{split}
\end{align}
where $\Delta$ is the shift of the fermion density onto the $b$ and $c$ sublattices, and
\begin{align}
A_{0}^a = \frac{j}{2\theta} + 4g\Delta , \quad A_{0}^b = A_{0}^c = -\frac{j}{2\theta} - 2g\Delta. \label{eqn:A0ansatz}
\end{align}
In our ansatz, by assuming the link currents, $j_k$, to be equal in magnitude and $n_b=n_c$, we have preserved as much of the lattice symmetry as possible (see Sec. \ref{sec:symmetries} for more details about  broken symmetries in our analysis).

For convenience we define 
\begin{align}
\Phi = (\Phi^a + \Phi^b + \Phi^c)/3 = 2\pi\frac{p}{q}= \frac{n_L}{3\theta}
\end{align}
to be the average statistical flux per unit plaquette where $p$ and $q$ are co-prime integers. Then one gauge choice which gives the flux configuration mandated by our mean field ansatz is
\begin{align}
\begin{split}
A_1({\bm x}) &= 0, \,\, A_2({\bm x}) = \Phi +\Delta / \theta, \,\,\,\,\qquad\quad A_3({\bm x}) = 0 \\
A_4({\bm x}) &= 0, \,\, A_5({\bm x}) = 3\Phi x_1 - \Phi - \Delta/\theta, \,\, A_6({\bm x}) = 3\Phi x_1
\end{split} \label{eqn:Akansatz}
\end{align}
where $x_i$ is the coordinate along the ${\bm e}_i$ direction. So at the mean-field level the problem reduces to one of fermions subject to constant magnetic and statistical fluxes and a staggered potential, the latter two of which must be solved for self-consistently, as described by the hopping Hamiltonian
\begin{widetext}
\begin{align}
\begin{split}
H(A) &= J \sum_{\bm x} ( \conj{\gamma}_- e^{iA_1({\bm x})}\ket{a,{\bm x}}\bra{b,{\bm x}} + \gamma_+ e^{iA_2({\bm x})}\ket{a,{\bm x}} \bra{c,{\bm x}} + \gamma_+ e^{iA_4({\bm x})}\ket{b,{\bm x}}\bra{a,{\bm x}+{\bm e}_1} + \conj{\gamma}_+ e^{iA_3({\bm x})}\ket{b,{\bm x}} \bra{c,{\bm x}+{\bm e}_1} \\ 
&+ \conj{\gamma}_- e^{iA_5({\bm x})}\ket{c,{\bm x}} \bra{a,{\bm x}+{\bm e}_2} + \gamma_- e^{iA_6({\bm x})}\ket{c,{\bm x}} \bra{b,{\bm x}+{\bm e}_2} + h.c.) + \sum_{{\bm x},\alpha} A_0^\alpha \ket{\alpha,{\bm x}}\bra{\alpha,{\bm x}}
\end{split}
\end{align}
\end{widetext}
where $\gamma_\pm = e^{i \phi_\pm / 3}$ are the phases arising from the background magnetic field and $\conj{\gamma}_\pm$ denote their complex conjugates. 

Now, as is explained in the previous section, the interaction term in the action is (formally) irrelevant as it involves more derivatives than the Chern-Simons term. So, as is done in the analysis of the continuum FQHE, we look for gapped CF states of the kinetic term. Formally this amounts to setting $g=0$. Note that this does not mean that the states we find can exist in the absence of interactions. As in the analysis of the continuum FQHE, there is a residual effect of the interactions in the composite fermion mapping which is what attaches the fluxes to the fermions. Additionally, at the mean field level, the interactions for $g\neq 0$ do affect the local energetics and and so can affect the sublattice imbalance $\Delta$. However, the states that we find in the (formal) $g=0$ limit should persist for finite $g\neq0$. Indeed, we show explicitly in Appendix \ref{sec:MFT_interactions} that the gapped states we find at a few sample fillings persist for finite interaction strength.

As the the filling of the lowest band of the original fermions $n_L$ increases, so too will the statistical flux. Examining the resulting Hofstadter spectrum as a function of filling will allow us to identify gapped states of the CFs. For each gapped state we can then compute the Chern number of the filled composite fermions bands which is given explicitly by \cite{Thouless82}
\begin{align}
C = \frac{1}{2\pi} \sum_{n\,\mathrm{filled}}\int_{\mathrm{BZ}} d^2 {\bm k} \, \mathcal{F}^n_{12} ({\bm k})
\end{align}
where the integral is performed over the Brillouin zone and $ \mathcal{F}^n_{12} ({\bm k}) = \epsilon_{ij} \partial_{{\bm k}_i}\mathcal{A}_j$ is the Berry curvature of the $n^\mathrm{th}$ band. The Berry connection is defined as $\mathcal{A}_j = -i\bra{n,{\bm k}} \partial_{{\bm k}_j} \ket{n,{\bm k}}$ where $\ket{n,{\bm k}}$ is the eigenvector in the $n^\mathrm{th}$ band of the Bloch Hamiltonian. We compute the Chern number numerically using the method of Ref. [\onlinecite{Fukui05}]. As discussed in the following section the Chern number appears in the effective topological theory of the FCI states.

We have plotted the Hofstadter spectrum, sublattice imbalance, and link currents as a function of in Fig. \ref{fig:k1s1_1pi2-a} for the case of $k=1$. For now we simply note that we find a number of sequences of gapped states, with the gapped states highlighted by the vertical red and purple lines occurring at fillings corresponding to the Jain sequence: $n_L = p / (2p+1), \, p \in \mathbb{Z}$. In the following section we will discuss the topological field theory describing these states and the pattern of Hall conductances exhibited by them.

\subsection{Symmetries and Mean Field Theory}
\label{sec:symmetries}
At this point we return to the issue of the explicit lattice point-group symmetry breaking of the action which is made manifest by this mean-field analysis. In a gapped insulating phase, although the net current must vanish, the current on a link, $\la j_k({\bm x}) \ra$, need not be zero. Indeed, in a generic chiral phase we expect to find currents circulating around the plaquettes. Eq. (\ref{eqn:MF2}) implies that non-zero currents require a staggered $A^\alpha_0 ({\bm x},t)$ (i.e. a spatially modulated Chern-Simons electric field). This staggered sublattice potential will create a staggered density of fermions. Hence our mean-field analysis would suggest that in general a ground state of CFs must necessarily break the point-group symmetry of the lattice (note that previous applications of this lattice Chern-Simons formalism, such as Ref. [\onlinecite{Kumar14}], did not self-consistently calculate the currents and so incorrectly found uniform states). The mean-field ansatz we have chosen preserves as much of the lattice symmetry as possible. This situation is to be contrasted with the CF theory of the continuum FQHE in which the mean-field ground state consists of fully filled CF Landau levels which have an exactly vanishing local current and hence a vanishing Chern-Simons electric field. Likewise, if we were to instead consider the problem of a square lattice in a uniform magnetic field then it would be possible to find a translationally invariant solution since such a state at the mean field level would have equal fluxes (the sum of the magnetic and statistical fluxes) through all plaquettes; it can be seen that the link currents must vanish in this state and hence the $A_0$'s would also vanish as per the self-consistent equations. As a result, the mean-field analysis reduces to a simple Hofstadter problem, as was discussed in Ref.~[\onlinecite{Moller15}], without the need to self-consistently solve for the currents and density imbalance.

Although numerical studies have predicted FCI states which spontaneously break lattice symmetries, \cite{Kourtis14} we suspect that the symmetry breaking in our analysis is an artifact of the mean field theory. 
In particular, note that in the $g=0$ limit we considered, although our ansatz with $n_a \neq n_b = n_c$ satisfies the mean field equations, configurations corresponding to rotations of this ansatz (e.g. with $n_b \neq n_a = n_c$) are not solutions
as is clear from Eq. (\ref{eqn:A0ansatz}) (which follows from only assuming that the link currents, $j_k$, are equal in magnitude). Hence there should not be an additional, trivial ground state degeneracy associated with this breaking of the point-group symmetry which suggests that this symmetry breaking is an artifact of the form of the lattice CS action. Furthermore, since the mapping of the fermions to CFs is exact at the level of the path integral, it should follow that all the symmetries of the original problem should be respected under the CF mapping at the full quantum level. Assuming that the effects of these corrections are not so large as to close the gap, the topological data we compute will be accurate. Since this is the focus of our study we will henceforth not concern ourself with the role played by the point-group symmetries, leaving a full analysis to future work.
Evidently an improvement over the saddle-point approximation is needed to correctly describe the non-topological properties of the candidate FCI states predicted using our Chern-Simons theory.

\section{Fractional Chern Insulator States}
\subsection{Topological Field Theory}
\label{sec:TQFT}

In the cases where the CFs are gapped, we can integrate them out to obtain an effective low-energy, continuum theory. Doing so will yield a Chern-Simons term with coefficient equal to the Chern number, $C$, of the filled CFs bands.\cite{Thouless82} Hence the effective continuum Lagrangian for fluctuations of the gauge fields about the mean field state is given by
\begin{equation}
\label{eqn: effective action}
\begin{split}
\mathcal{L}_{\mathrm{eff}}[A_\mu,B_\mu] =  & \frac{C}{4\pi} \epsilon^{\mu\nu\lambda} A_\mu \partial_\nu A_\lambda + \mathcal{L}_{CS}^{\mathrm{ctm}}[A_\mu, B_\mu] + \dots
\end{split}
\end{equation}
where $\ldots$ represent irrelevant terms.

Adding in quasi-particle currents with gauge charges $l_I$ and a coupling to an external probe gauge field $\tilde{A}_\mu$, we can write our theory in the conventional form:
\begin{align}
\mathcal{L} = \frac{K_{IJ}}{4\pi}\epsilon^{\mu\nu\lambda} a_\mu^I \partial_\nu a_\lambda^J - \frac{q_I}{2\pi} \epsilon^{\mu\nu\lambda} \tilde{A}_\mu \partial_\nu a^I_\lambda + l_I j^\mu_{\mathrm{qp}} a_\mu^I \label{eqn:KMatrixCtmAction}
\end{align}
where\footnote{This theory is equivalent to that arising from the standard hierarchical construction.\cite{Wen92,Wen95} In particular, the Hall conductance and anyon content of the theories are readily checked to be identical.}
\begin{align}
K_{IJ} =  \begin{pmatrix}
  -2k & 1 & 0 \\
  1   & C & 0 \\
  0   & 0 & 1
 \end{pmatrix}, \,\,
q_I = \begin{pmatrix}
  1 \\
  0 \\
  0  
 \end{pmatrix}, \,\,
a^\mu_I = \begin{pmatrix}
  B^\mu \\
  A^\mu \\
  D^\mu  
 \end{pmatrix}.
\end{align}
Here, we introduced the gauge field $D^\mu$ as the bare quasi-particles are CFs and so possess fermionic statistics. To account for this, and because the quasi-particles do not couple to $B^\mu$, the flux vector is restricted to have form ${\bm l} = (0, l, l)^T, \, l \in \mathbb{Z}$.\cite{Lopez99} Integrating out the Chern-Simons fields we find the Hall conductance to be (in units of $e^2/h$)
\begin{align}
\sigma_{xy} = -  {\bm q}^T K^{-1} {\bm q} =  \frac{C}{2kC + 1}. \label{eqn:HallCond}
\end{align}
Likewise, the quasi-particle charges and statistics are
\begin{align}
Q_{\bm l} = -{\bm l}^T K^{-1} {\bm q}, \quad \theta_{{\bm l}{\bm l}'} = -2\pi {\bm l}^T K^{-1} {\bm l}'.
\end{align}
The ground state degeneracy $g$ on a torus is then\cite{Wen95}
\begin{equation}
g=|\mathrm{det} \, K| = |2kC+1|.
\end{equation}

We can also extract the modular properties of the theory from the effective, topological action. In particular, the components of the modular $\mathcal{S}$ and $\mathcal{T}$ matrices (both of rank $|\textrm{det}K|$, the number of anyons) are given by 
\begin{align}
\mathcal{S}_{ab} = \frac{1}{\sqrt{|\mathrm{det} K|}}e^{-2\pi i {\bm l}_a^T K^{-1} {\bm l}_b}, \quad \mathcal{T}_{aa} = e^{-\pi i {\bm l}_a^T K^{-1} {\bm l}_a}
\end{align}
where the total quantum dimension is $\mathcal{D} = \sqrt{|\mathrm{det} K|} = S_{00}^{-1}$, a topological invariant that determines the universal entanglement properties.\cite{Levin-2006,Kitaev-2006,Dong-2008}
Moreover, the full topological structure of theory is encoded in $\mathcal{S}$ and $\mathcal{T}$. Hence the topological field theory for an FCI state which can be described by a gapped state of CFs is wholly determined by $k$, the number of attached fluxes, and $C$, the Chern number of the filled CF bands.

\subsection{Fractionalized States from the CF Hofstadter Spectrum}
\label{sec:hofstadter}

Following the results of the previous sections, we now analyse the Hofstadter spectra of the CFs to identify candidate FCI states on the kagome lattice. Before inspecting the spectrum, we note that a gapped state of the CFs must satisfy the Diophantine equation 
\begin{equation}
n_L = -\frac{3\Phi}{2\pi} C + r, \, r \in \mathbb{Z}
\end{equation}
 where $C$ is the Chern number of the filled CF bands \cite{Moller15} and $-3\Phi$ is the net flux per unit cell (note that because of the coupling between the fermions and $A_\mu$, if the statistical flux through a plaquette is $\phi$, the fermions feel a flux $-\phi$). Combined with the flux attachment condition, this implies that a gapped state must satisfy
\begin{align}
n_L = \frac{r}{2kC+1} = \frac{r}{C}\sigma_{xy}
\end{align} 
The existence of states satisfying $r \neq C$ is made possible due to the presence of the lattice.

\begin{figure}%
  \subfigure[]{
     \includegraphics[width=0.45\textwidth]{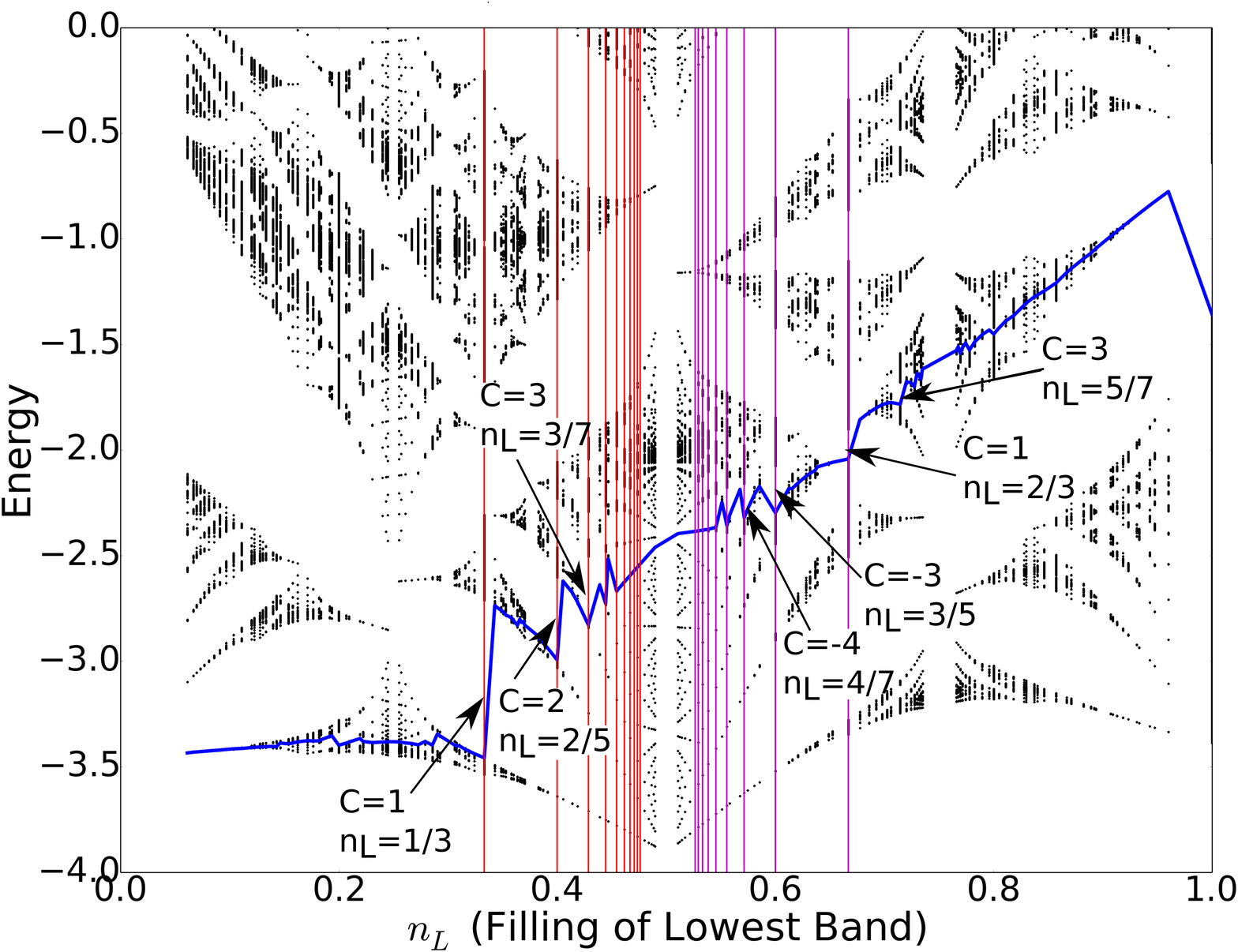} \label{fig:k1s1_1pi2-a}
  }
  \subfigure[]{
     \includegraphics[width=0.45\textwidth]{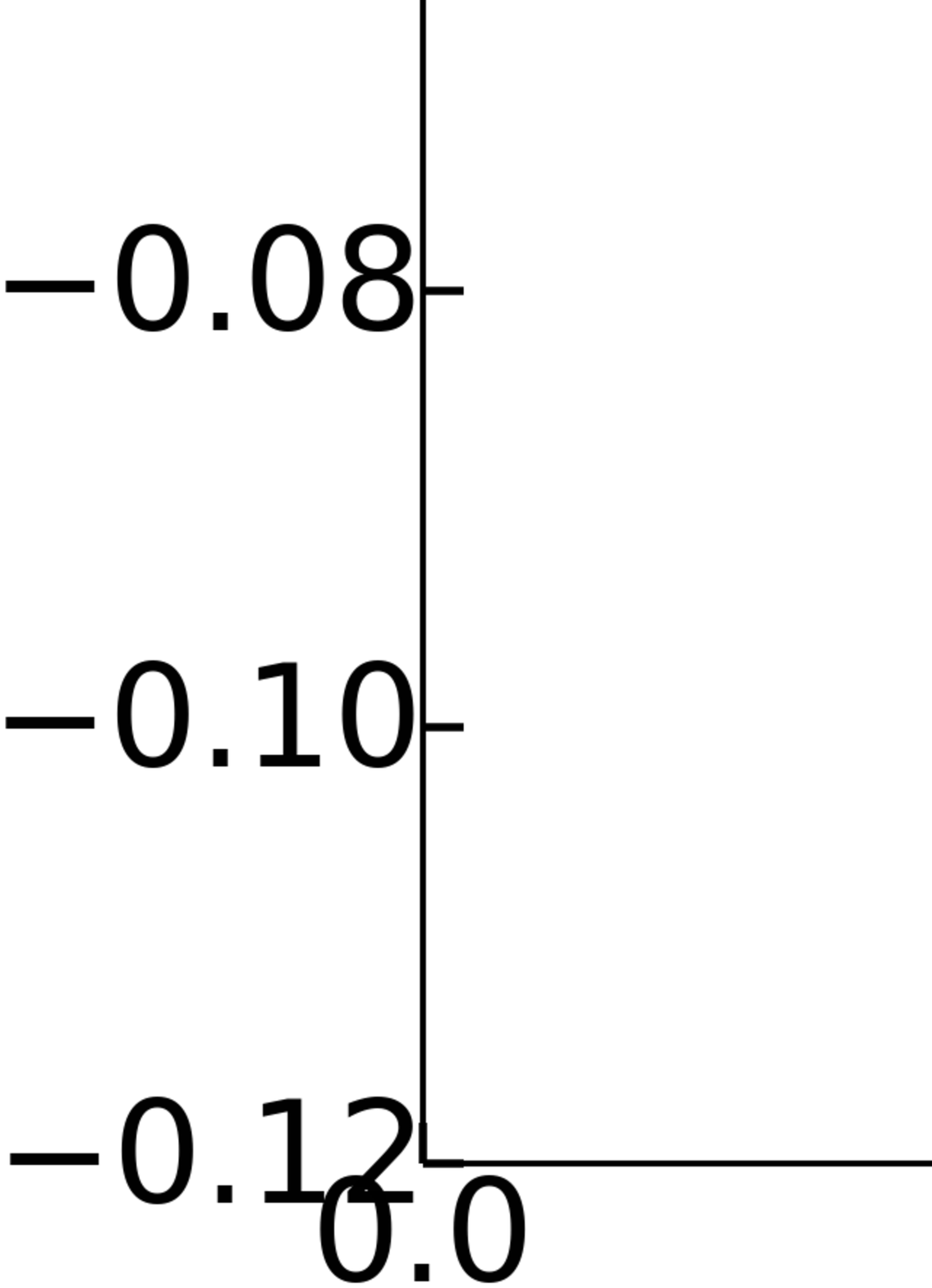} \label{fig:k1s1_1pi2-b}
  }
  \subfigure[]{
     \includegraphics[width=0.45\textwidth]{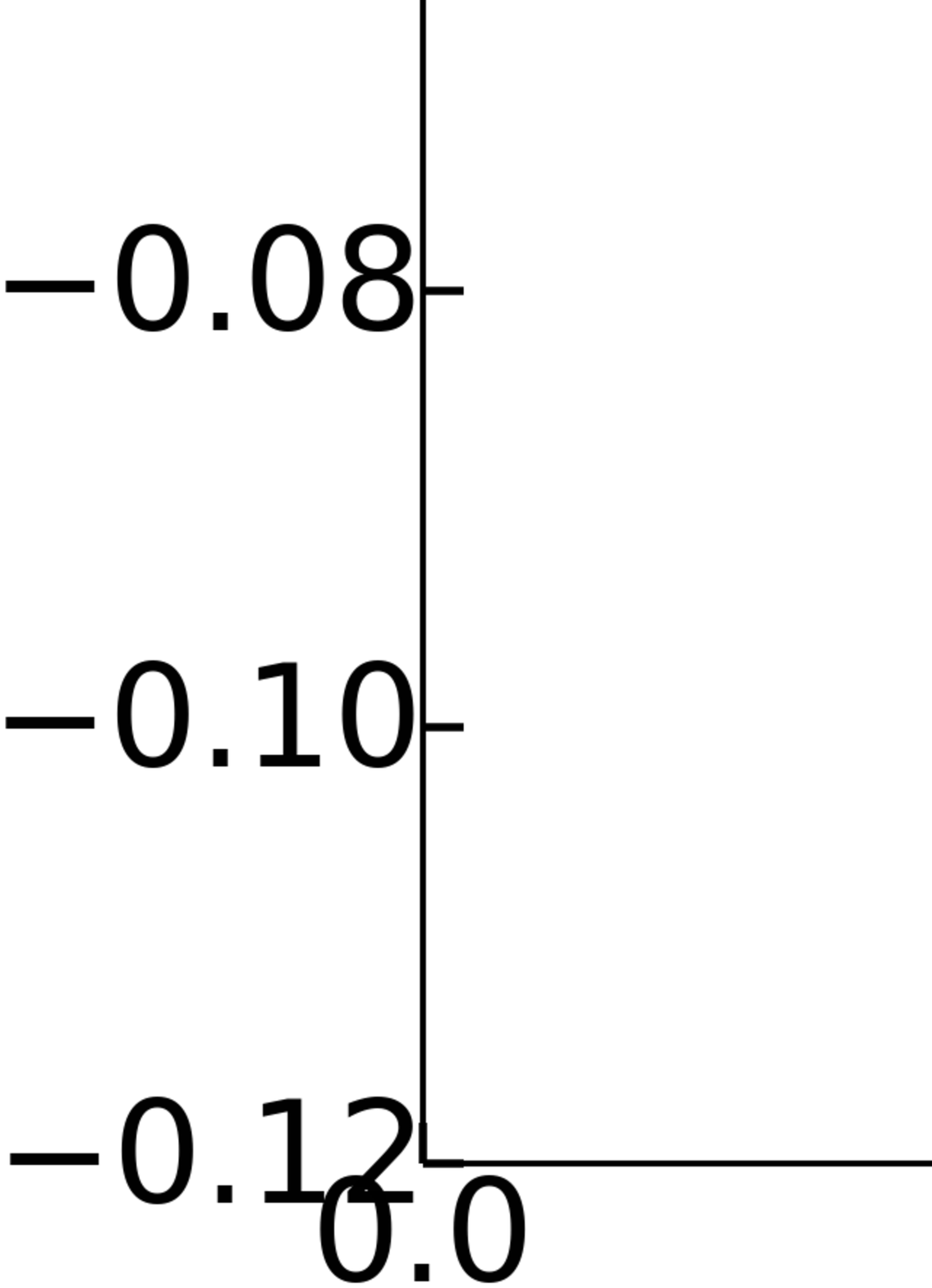} \label{fig:k1s1_1pi2-c}
  }
  \caption{(Color online) a) Composite fermion Hofstadter spectrum on the kagome lattice with $k=1$ and $\phi_\pm = \pi / 2$. The blue line is the Fermi energy. Some examples of gapped states are labelled with their filling and Hall conductance. Vertical red (purple) lines are drawn at fillings corresponding to the principal particle (hole) Jain sequence. The ratio of the mean field sublattice shift, $\Delta$, to the filling and the mean field current per link, $k$, are plotted in b) and c), respectively. All quantities are in units of $J=1$.}
  \label{fig:k1s1_1pi2}
\end{figure}

Turning to our  model, Fig. \ref{fig:k1s1_1pi2-a} depicts a portion of the Hofstadter spectrum on the kagome lattice and Fermi Energy for the case of $k=1$ pair of attached fluxes. Figs. \ref{fig:k1s1_1pi2-b}-(c) depict the mean field sublattice density shift, $\Delta$, and link current, $j$, as a function of filling. It is clear from the Hofstadter spectrum that gapped states exist at fillings given by the principal Jain sequence and the gap sizes approach zero as $n_L$ approaches $1/2$, as is the case in the continuum. We have also labelled the first few states in this sequence with the Chern number of the filled CFs bands and the filling factor. Using the expression for the Hall conductance given in Eq. \eqref{eqn:HallCond}, we find that $\sigma_{xy} = n_L$ (except for the state at $n_L=2/3$). Hence we recover the principal Jain sequence despite the absence of a net non-zero magnetic field. We have not extensively analysed the spectrum of mean field states for $k\neq 1$ but our preliminary numerics suggest that for $|k|>0$, there should generically exist gapped states of the CFs at filling factors corresponding to the Jain sequence of states 
$n_L = p/(2kp + 1),\, p\in \mathbb{Z} \backslash\{0\}$
 with Hall conductances satisfying $\sigma_{xy} = \mathrm{sgn}(k) n_L$. These states are  analogous to the  FQH Jain sequences.
We note, however, that there are exceptions to this rule. For instance, as noted above, the $k=1$ state at $n_L=2/3$ has $\sigma_{xy}=1/3\neq n_L$ and so is the not a typical Jain state.
In Appendix \ref{sec:UniformDensity} we present for comparison the Hofstadter spectrum obtained if one assumes a uniform density of fermions (and hence uniform statistical flux configuration) and does not correctly solve for the link current self-consistently via the mean field equations.

Now, in the conventional FQH, one can condense quasiparticles in a Jain state to form a FQH state with a filling fraction which does not lie in the principal Jain sequence. On the lattice, we instead find what is presumably a dense set of FCI states not lying in the Jain sequences without invoking this condensation mechanism. In particular, this means that there exist FCI states with the same Hall conductance but at different fillings. For instance, we find gapped states at $n_L = 1/7, 2/7, 3/7, 5/7$,  all with $\sigma_{xy} = 3/7$. The $n_L = 5/7$ state is highlighted in Fig. \ref{fig:k1s1_1pi2-a}. It should be noted, however, that interactions will likely render the lattice-specific states with the smallest gaps energetically unfavorable relative to topologically trivial phases (e.g. Wigner crystals, CDWs, nematic states), which we have not considered.

\section{Symmetry Fractionalization}
\label{sec:symmetry-fractionalization}

 The question now arises as to how states at different fillings but with the same topological order can be classified. In order to answer this question, we make use of the fact that topological phases enriched with symmetries (known as SETs) possess anyonic excitations which can transform projectively under symmetry operations.\cite{Chen17} This phenomenon is known as symmetry fractionalization and implies that anyons can carry fractional symmetry charges. In particular, SETs can be distinguished by their symmetry fractionalization class (i.e. the set of projective phases for each anyon). Given an SET with Abelian anyon group $\mathcal{A}$ and symmetry group $\mathcal{G}$, the set of distinct symmetry fractionalization classes is given by the second cohomology group $H^2[\mathcal{G},\mathcal{A}]$.\cite{Essin13,Barkeshli14}

Now, the anyon group of our FCI states is $\mathcal{A} = \mathbb{Z}_{m}$, where $m=|\mathrm{det} K|$, and the symmetry group is $\mathcal{G} = \mathbb{Z}^2 \times U(1)$ arising from lattice translation symmetry and $U(1)$ charge conservation, respectively.\footnote{The kagome lattice has a larger space group symmetry but, for simplicity, we will focus on the symmetry group $\mathcal{G} = \mathbb{Z}^2 \times U(1)$.} Hence, the distinct fractionalization classes are given by $H^2[\mathbb{Z}^2 \times U(1),\mathbb{Z}_{m}] = \mathbb{Z}_m \times \mathbb{Z}_m$. The fractionalization is given on specification of the fluxon/vison $F$ and the background anyon $\mathfrak{b}$.\cite{Cheng16} Physically speaking, the fluxon is the anyon created on the insertion of a $2\pi$ flux quantum; such an excitation carries charge equal to $\sigma_{xy}$. The fluxon specifies the $U(1)$ fractionalization in that the charge of an anyon, $Q_a$, is determined via the mutual statistics between $a$ and $F$: $\exp(2\pi i Q_a) = \exp(i\theta_{F,a})$. Similarly, the background anyon specifies the translational symmetry fractionalization. This anyon possesses charge equal to the charge density per unit cell, $n_L$, and so physically one can view the ground state as a crystal of the background anyon $\mathfrak{b}$, with one $\mathfrak{b}$ residing in each unit cell.\cite{Cheng16,Lu17} Braiding an anyon $a$ around a single unit cell will give a phase $\exp(i \theta_{a,\mathfrak{b}})$ which implies
\begin{align}
(T^a_2)^{-1} (T^a_1)^{-1} T^a_2 T^a_1 = e^{i \theta_{a,\mathfrak{b}}}
\end{align}
where $T^a_{1,2}$ are the local translations along the $e_{1,2}$ directions acting on anyon $a$. 
As an aside, we note that the fact that the system realizes a projective representation of the translation symmetry group may be viewed as an quantum anomaly of the discrete translational symmetry. Hence this effect may also lead to momentum pumping on a torus with tilted boundary conditions, a phenomenon which may be interesting to study in future work.

This analysis provides an interesting perspective on our spectrum of FCI states. The Jain states satisfy $\sigma_{xy} = n_L$ and so realize the fractionalization classes for which $F = \mathfrak{b}$. However, given $\mathcal{A} = \mathbb{Z}_{|\mathrm{det} K|}$ and the fluxon $F$ (equivalently, $\sigma_{xy}$) there are a total of $|\mathrm{det} K|$ translational fractionalization patterns which can be realized from the choice of background anyon $\mathfrak{b}$. We have shown that these other fractionalization classes which have $\mathfrak{b} \neq F$ (i.e. $n_L \neq \sigma_{xy}$) can be realized on the lattice.

Moreover, we can use this language to make  statements about the momenta of the topologically degenerate ground states.\cite{Essin13} Consider a gapped FCI state with $|\mathrm{det} K| = m$ and background anyon $\mathfrak{b}$ on a torus of size $N_1 N_2$ where $N_{1,2}$ are the number of unit cells in the $e_{1,2}$ direction, and $N_2$ is co-prime with $m$. Now, let $\ket{0}$ be the  ground state labeled by the trivial anyon, and let $\ket{\phi},\dots,\ket{\phi^{m-1}}$ be the ground states generated by applying Wilson loop operators, $\mathcal{L}_2^{\phi^n}$, to $\ket{0}$ where $\phi$ is the minimal charge anyon. The Wilson loops  can be viewed as operators creating  anyon-antianyon pairs, braiding the anyons around a cycle of the torus, and then fusing the anyon with the anti-anyon. We can thus make the identification 
\begin{equation}
\mathcal{L}_\mu^a = (T_\mu^a)^{N_\mu}
\end{equation}
 where $\mu = 1,2$. This implies that
\begin{equation}
T_1 \mathcal{L}_2^a T_1^{-1} = T_1 (T^a_2)^{N_2} T_1^{-1} = e^{-i N_2 \theta_{a,\mathfrak{b}}} \mathcal{L}_2^a.
\end{equation}
 Now, we have that $\ket{\phi^n} = (\mathcal{L}_2^\phi)^n \ket{0}$. Without loss of generality, suppose that the trivial ground state $\ket{0}$ has zero momentum. Hence,
\begin{align}
T_1 \ket{\phi^n} = e^{-i N_2 n \theta_{\phi,\mathfrak{b}}} \ket{\phi^n}.
\end{align} 
So, relative to the trivial state $\ket{0}$, the states $\ket{\phi^n}$ will have momenta ${\bm k}_n = (N_2 n \theta_{\phi,\mathfrak{b}},0)$. Since this momentum shift depends on the braiding angle with the background anyon, it provides a clear  way to distinguish between two FCIs at different fillings possessing the same topological order. 

As an explicit example, consider an FCI state with $\sigma_{xy} = C/(2kC+1)$ and $n_L = r/(2kC+1)$. In terms of the quasiparticle vectors ${\bm l}$, it is readily seen that the fluxon is represented by ${\bm l}_F = -(0,C,C)^T$, the background anyon by ${\bm l}_\mathfrak{b} = -(0,r,r)^T$, and the minimal charge anyon by ${\bm l}_{\phi} = -(0,1,1)^T$. The translational symmetry fractionalization for the minimal anyon is then obtained by using the fact that 
\begin{equation}
\theta_{{\bm l}_\mathfrak{b}, {\bm l}_\phi} = -2rk\frac{2\pi}{2kC+1}.
\end{equation}
 We have listed the fractionalization patterns for observed FCI states with $\sigma_{xy} = 3/7$ in Table \ref{tab:SymmFrac}.

\begin{table}[hbt]
\caption{\label{tab:SymmFrac}%
Symmetry fractionalization for FCI states with $\sigma_{xy} = 3/7$. The third column gives the translational symmetry fractionalization for the minimal charge anyon $\phi$. The fourth column gives the $e_1$ component of the crystal momentum of the ground state $\ket{\phi}$ relative to the trivial ground state.
}
\begin{ruledtabular}
\begin{tabular}{cccc}
\textrm{$n_L$}&
\textrm{$\sigma_{xy}$}&
\textrm{$(T^\phi_2)^{-1} (T^\phi_1)^{-1} T^\phi_2 T^\phi_1$}&
\textrm{$\hat{k}_1\ket{\phi}$}\\
\colrule
$1/7$ & $3/7$ & $\exp\left( -2 \frac{2\pi}{7} i \right)$  & $-2 \frac{2\pi}{7}N_2$ \\
$2/7$ & $3/7$ & $\exp\left( -4 \frac{2\pi}{7} i \right)$  & $-4 \frac{2\pi}{7}N_2$ \\
$3/7$ & $3/7$ & $\exp\left( -6 \frac{2\pi}{7} i \right)$  & $-6 \frac{2\pi}{7}N_2$ \\
$5/7$ & $3/7$ & $\exp\left( -10 \frac{2\pi}{7} i \right)$ & $-10 \frac{2\pi}{7}N_2$ \\
\end{tabular}
\end{ruledtabular}
\end{table}

\section{Distinction between FCIs and the Lattice FQHE}
\label{sec:FCI vs FQHE}

Lastly, we would like to emphasize that the system we studied has a net zero external magnetic field and it is in this sense that the fractionalized states we find should be called FCI states. Conversely, fractionalized states found in lattice systems subject to a uniform magnetic field (i.e. in Hofstadter bands) should be considered lattice FQH states. Although both exhibit similar physics, it is important to make clear the distinction that one requires a net non-zero magnetic field while the other does not. In that regard, the states observed in Ref. [\onlinecite{Spanton17}] are lattice FQH states. The experimental observation of an FCI remains an open problem.

\section{Conclusions}
\label{sec:conclusions}

We  formulated a composite fermion theory of Fractional Chern Insulator states on the kagome lattice using a consistent lattice Chern-Simons theory. We find that partial filling of the lowest band yields two types of sequences of gapped states: those which satisfy $\sigma_{xy} = n_L$ and those which do not. Hence our theory provides a series of candidate FCI states whose stability against local interactions may be tested in numerical and experimental studies. Using the language of SETs we illustrated how these states may be viewed as realizing distinct symmetry fractionalization classes which exposes the rich structure of FCI states and allows for concrete, numerically verifiable, statements about ground state quantum numbers.

\begin{acknowledgments}
The authors thank Z. Liu and N. Regnault for useful discussions and sharing their unpublished work. This work was supported in part by National Science Foundation grants DMR 1408713 and DMR 1725401 at the University of Illinois (RS, EF) and  by the Gordon and Betty Moore Foundation EPiQS Initiative through Grant No. GBMF4305 (LS).
\end{acknowledgments}

\appendix

\section{Non-Interacting Model Band Structure}
\label{sec:non-interacting band stucture}
\begin{figure}%
   \includegraphics[width=0.48\textwidth]{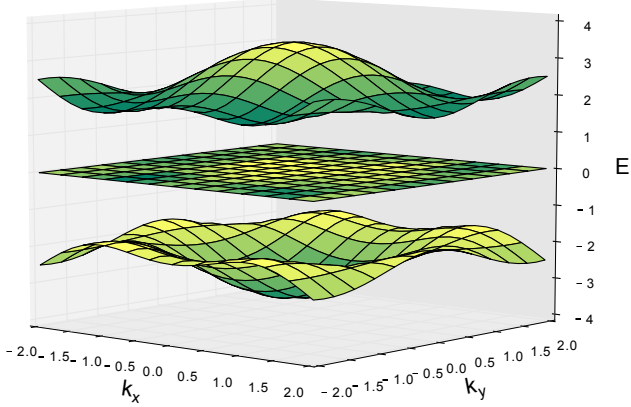}
  \caption{(Color online) Band structure of the model given by Eq. (\ref{eqn:model}) in the absence of interactions with $\phi_\pm = \pi/2$ and $J=1$. The lower, middle, and upper bands have $C_0=+1,0,-1$, respectively}
  \label{fig:non-interacting-spectrum}
\end{figure}%
In Fig. \ref{fig:non-interacting-spectrum} we have plotted the band structure of the model Eq. (\ref{eqn:model}) for $g=0$, $\phi_\pm = \pi/2$, and $J=1$. The lowest band, which we partially fill, has Chern number $C_0=+1$. Note that the bandwidth of this band and the band gap are comparable in magnitude (the former is slightly smaller than the latter).

\section{Comparison with Uniform Density Approximation}
\label{sec:UniformDensity}
As noted in the main text our lattice Chern-Simons action explicitly breaks the lattice symmetries and as a consequence the mean-field ground state the theory predicts on the Kagome lattice breaks them as well. Previous applications of this method (e.g. the chiral spin liquid study of Ref. [\onlinecite{Kumar14}]) did not correctly solve the mean field equations (they assumed the currents to be zero) and so found uniform states. To reiterate what is said in the main text, we believe that at the full quantum level the lattice symmetries broken by our mean field solution should be restored. However this is a nontrivial calculation (presumably being a non-perturbative effect) so it is worthwhile to compare the results of our mean-field theory with those one would obtain if one assumed a uniform ground state with equal statistical fluxes through all plaquettes (which we stress is not a valid solution to the mean field equations of our theory).

\begin{figure}%
   \includegraphics[width=0.48\textwidth]{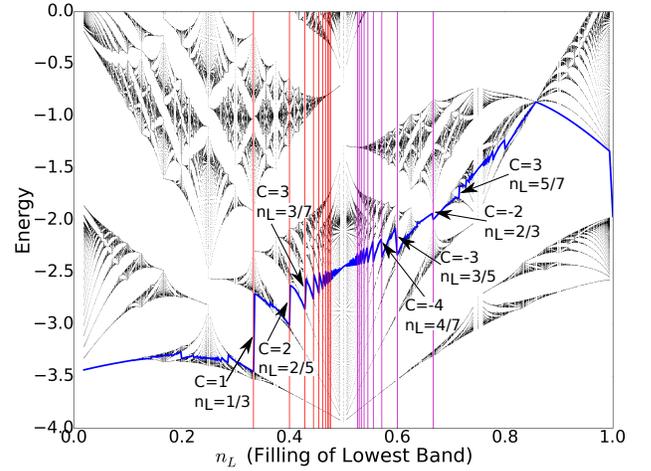}
  \caption{(Color online) Composite fermion Hofstadter spectrum on the kagome lattice with $k=1$ and $\phi_\pm = \pi / 2$ assuming a uniform density of composite fermions and equal fluxes through all plaquettes. The blue line is the Fermi energy. Some examples of gapped states are labelled with their filling and Hall conductance. Vertical red (purple) lines are drawn at fillings corresponding to the principal particle (hole) Jain sequence.}
  \label{fig:k1s1_1pi2_uniform}
\end{figure}%

In Fig. \ref{fig:k1s1_1pi2_uniform} we have plotted the Hofstadter spectrum for this uniform flux approximation. Note that the spectrum here has much finer detail than Fig. \ref{fig:k1s1_1pi2-a} of the main text as in this case we are not solving the mean field equations self-consistently but rather simply computing the band structure with the aforementioned uniform fluxes. We see that Fig. \ref{fig:k1s1_1pi2-a} and Fig. \ref{fig:k1s1_1pi2_uniform} are largely similar. In particular gapped states exist at the same fillings, including the Jain sequence. The Chern numbers are mostly unchanged for the fillings we have checked; one exception is the gapped state at $n_L = 2/3$ filling. Our mean-field analysis predicts this gapped state to have $\sigma_{xy} = 1/3$ while the uniform approximation would suggest a state with $\sigma_{xy} = 2/3$. The gap of this state in both schemes is small, however, so it is unlikely that it would survive competition with other ordered states.

\section{Full Self-Consistent Solution} \label{sec:MFT_interactions}
In this Appendix we repeat our mean field analysis for $g\neq 0$ to check the stability of the gapped states we predict. As explained in Sec. \ref{sec: Mean Field Theory}, although the situation considered in the main text corresponds to formally taking $g=0$, this does not mean that we neglected the role of interactions). Now, in our mean field analysis we found that even for $g = 0$, the sublattice imbalance, $\Delta$, is generically non-zero, presumably as a result of the explicit point-group symmetry breaking of our lattice Chern-Simons action. In the case of $g\neq0$ there is a finite energy cost associated with this imbalance and thus any finite value for the interaction strength, $g$, will affect the values of the sublattice densities. However, provided $g$ is smaller than some critical $g_c$, we expect in general that the sublattice density will vary continuously and slowly so that the gapped states predicted in the main text remain gapped. In order to illustrate this, we perform a mean field analysis of our theory with the interaction term, Eq. (\ref{eqn:S_int}), where $V_{\alpha\beta}(\mathbf{x} - \mathbf{y}) = 1$ if $(\mathbf{x},\alpha)$ and $(\mathbf{y},\beta)$ are nearest neighbors and $V_{\alpha\beta}(\mathbf{x} - \mathbf{y}) = 0$ otherwise. As before we focus on time independent solutions of Eq. (\ref{eqn:MF1}) and Eq. (\ref{eqn:MF2}) which preserve translational symmetry. Using these assumptions we note that we can re-write Eq. (\ref{eqn:MF2}) as
\begin{align}
\begin{split}
\la j_k \ra &= -\theta (-1)^k (A_0^a - f_k A_0^c - (1-f_k) A_0^c ) \\ 
& - 2 g \theta^2 (-1)^k \left(\Phi^a - f_k \Phi^c - (1-f_k) \Phi^b \right)
\end{split}
\end{align}
where $f_k = 1$ for $k = 1,5,6$ and $f_k = 0$ for $k = 2,3,4$. 

\begin{figure*}
  \subfigure[\; $n_L = 1/3$]{
     \includegraphics[width=0.4\textwidth]{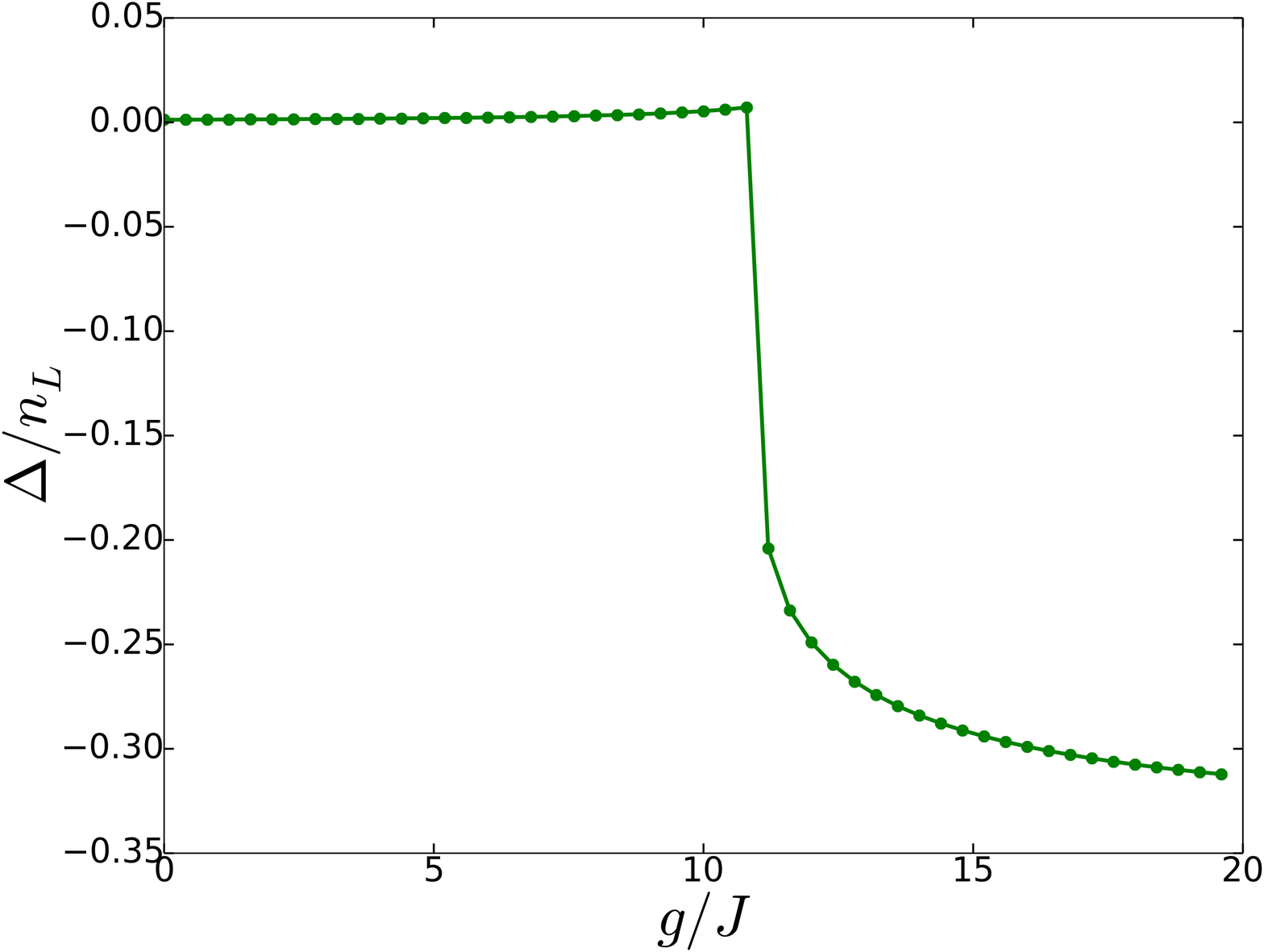}
  }
  \subfigure[\; $n_L = 1/3$]{
     \includegraphics[width=0.4\textwidth]{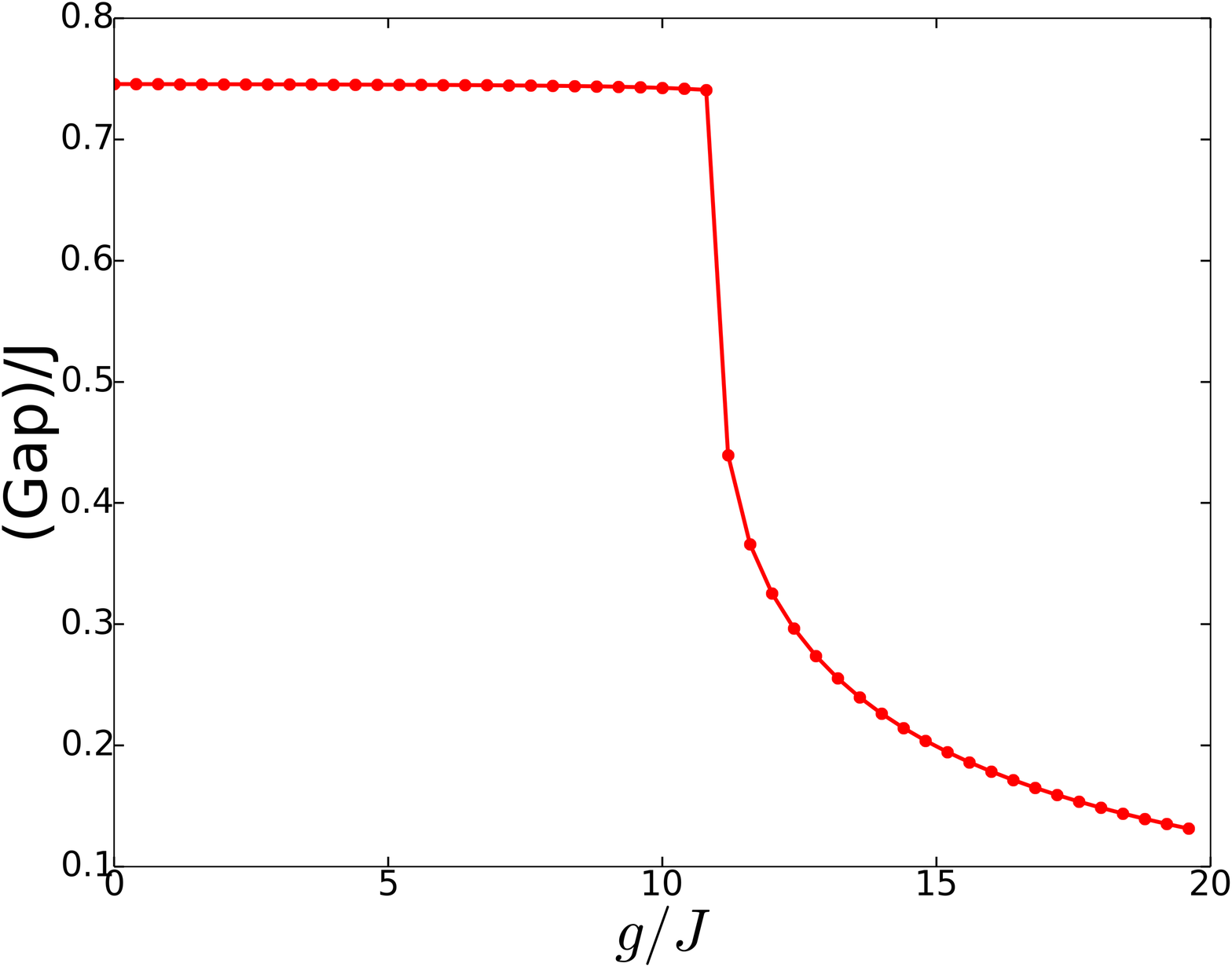}
  }
  \subfigure[\; $n_L = 2/3$]{
     \includegraphics[width=0.4\textwidth]{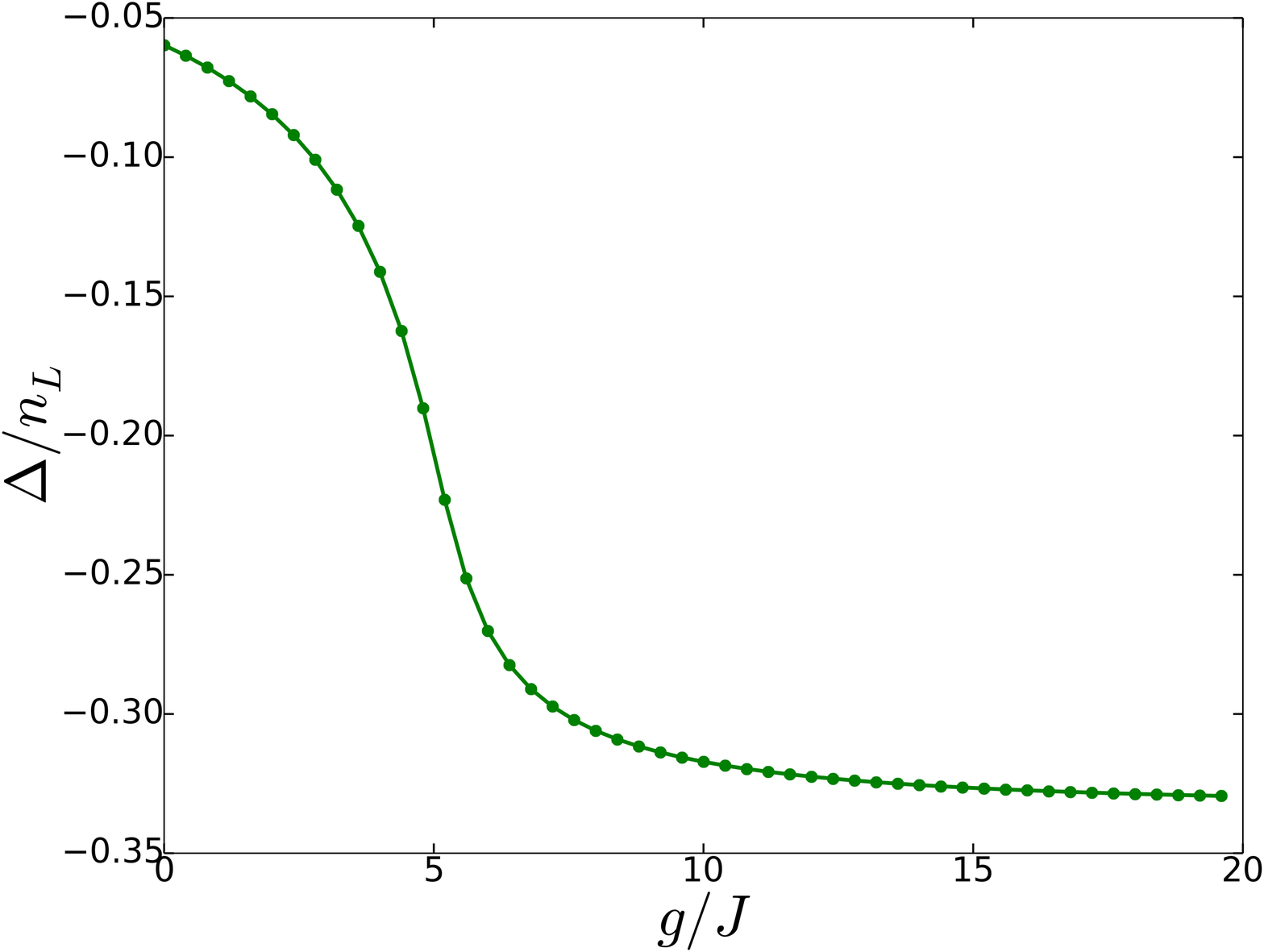}
  }
  \subfigure[\; $n_L = 2/3$]{
     \includegraphics[width=0.4\textwidth]{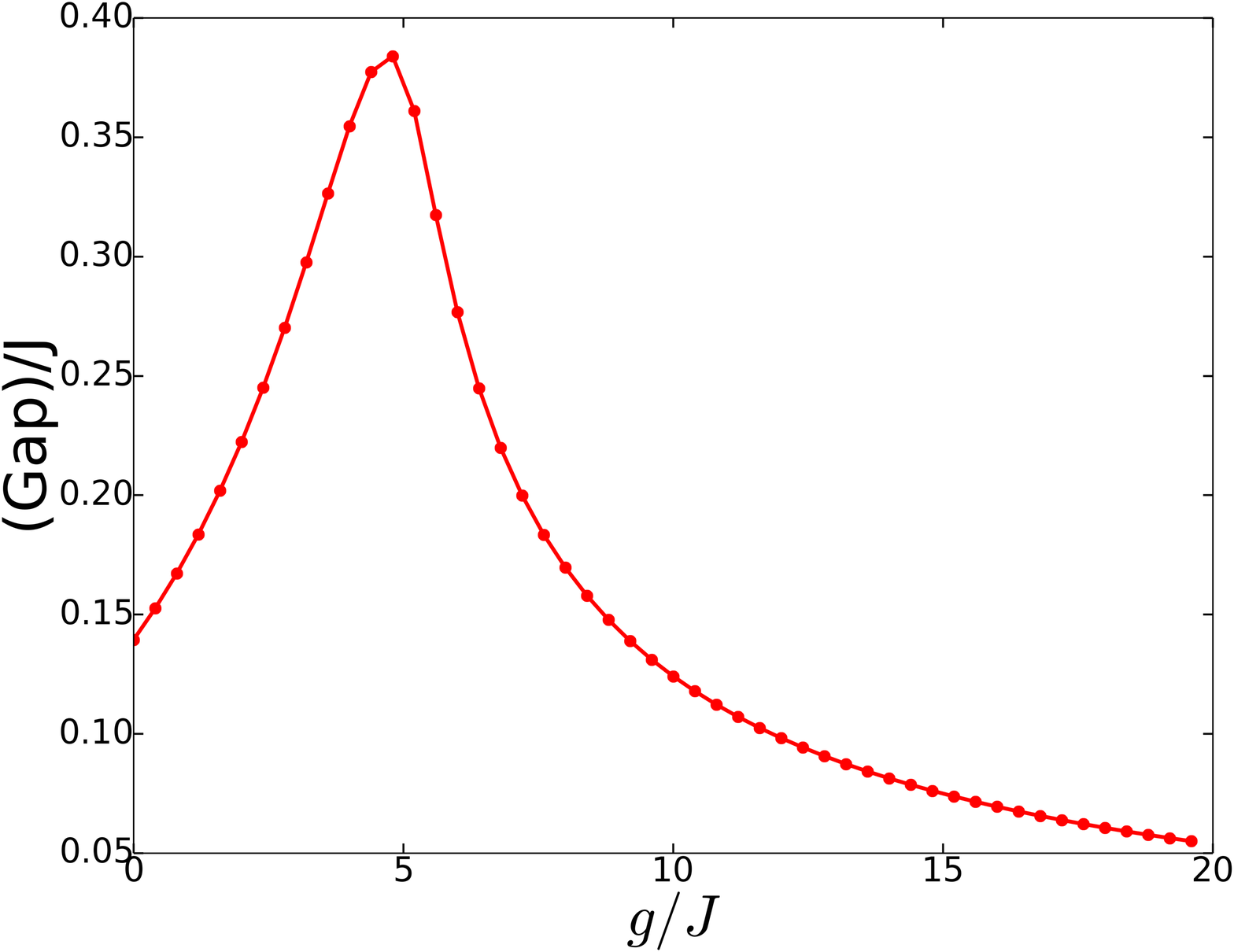}
  }
  \caption{(Color online) Plots of the sublattice imbalance and band gap for (a,b) $n_L = 1/3$ and (c,d) $n_L =2/3$ as a function of interaction strength $g$. Note that for small $g$ the imbalance, $\Delta$, and band gap vary smoothly with the latter never vanishing.} \label{fig:MFT_interactions}
\end{figure*}
For simplicity we have focused on the cases of $n_L = 1/3$ and $n_L = 2/3$. In Fig. \ref{fig:MFT_interactions} we have plotted $\Delta$ and the band gap as a function of $g$ at these two fillings.
It is clear in the case of $n_L = 1/3$ that the band gap does not close and $\Delta$ varies smoothly up to a critical value of $g$.
Likewise in the case of $n_L = 2/3$ the imbalance $\Delta$ varies smoothly. 
The jump of $\Delta$ in the case of $n_L=1/3$ appears to signal a phase transition to a nematic state. However, as discussed in the main text, since our Chern-Simons lattice action explicitly breaks the point-group symmetry we cannot trust our mean-field analysis to make accurate predictions about spontaneous rotational symmetry breaking.
Nevertheless, this data suggests that we are justified in assuming that small, finite interactions will not affect the topological properties of the states predicted by our mean field analysis.

\section{Spectrum of the $\mathcal{M}$-Matrix}
\label{sec:spectrum of K}
Proper implementation of the lattice Chern-Simons theory requires that the matrix kernel Eq.\eqref{eqn: kagome matrix kernel} be
non-singular, so as to guarantee that the commutation relations $[A_i({\bm x}),A_j({\bm y})] = -\frac{i}{\theta}\mathcal{M}^{-1}_{ij}({\bm x}-{\bm y})$ are well defined.
To access the eigenvalues of the $\mathcal{M}_{ij}(\bf{x} - \bf{y})$, we work with its Fourier transform $\mathcal{M}_{ij}(\bf{q})$ obtained
by substituting the displacement operators $s_{j}$, $j=1,2$, by their Fourier representation 
$s_{j}({\bf{q}}) = e^{-i q_{j}}$, where $q_{j} = {\bf{q}}\cdot{\bf{e}}_{j}$ is the momentum component
along the direction defined by the unit vector ${\bf{e}}_{j}$. With that, $\mathcal{M}({\bf{q}})$ is seen to be an anti-Hermitian matrix. Then $i\,\mathcal{M}({\bf{q}})$ is a Hermitian $6 \times 6$ matrix, whose eigenvalues are found to be non-zero, hence $\mathcal{M}$ is invertible.
To illustrate the non-singular character of the matrix kernel, we plot below the eigenvalues of $i\,\mathcal{M}({\bf{q}})$ as function of $q_{1}$ for the choice $q_{2} = \pi$.

\begin{figure}[hbt]
  \includegraphics[width=0.37\textwidth]{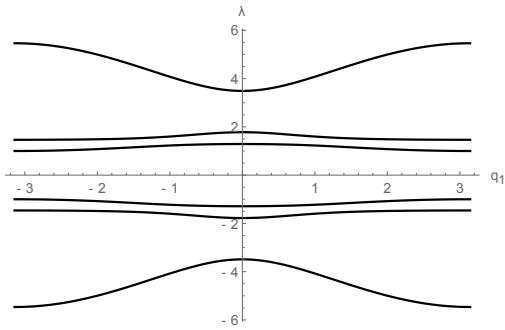}
  \caption{(Color online) Spectrum of the Hermitian $6 \times 6$ matrix $i\,\mathcal{M}({\bf{q}})$ as function of $q_{1}$ for the choice $q_{2} = \pi$.}
  \label{fig:K-matrix-spectrum}
\end{figure}

\providecommand{\noopsort}[1]{}\providecommand{\singleletter}[1]{#1}%
\end{document}